\documentclass[11pt, a4paper]{elsarticle}
\usepackage[top=30truemm, bottom=30truemm]{geometry}
\usepackage{newtxtext}

\usepackage{palatino} 

\usepackage{amsmath, amssymb, amsfonts, mathpazo} 
\usepackage{bm}
\usepackage{cases}

\usepackage[driverfallback=dvipdfmx]{hyperref}
\hypersetup{colorlinks=true}

\usepackage{graphicx}
\usepackage{natbib}

\biboptions{authoryear}
\usepackage{framed}
{\endMakeFramed}
\usepackage{mdframed}
\usepackage{amsthm} 

\usepackage{algpseudocode}
\usepackage{algorithm}

\usepackage{multirow,bigdelim}

\usepackage{listliketab}
\usepackage{booktabs}
\usepackage{tabularx}
\newcolumntype{C}{>{\centering\arraybackslash}X}
\newcolumntype{L}{>{\raggedright\arraybackslash}X}
\newcolumntype{R}{>{\raggedleft\arraybackslash}X}
\usepackage{longtable}

\usepackage{changes}
\definechangesauthor[color=red]{a}
\newcommand{\stkout}[1]{\ifmmode\text{\sout{\ensuremath{#1}}}\else\sout{#1}\fi}
\setdeletedmarkup{\stkout{#1}}

\journal{}
\begin{document}

\begin{frontmatter}

\title{\textbf{Capturing positive network attributes during the estimation of \\recursive logit models: A prism-based approach}}

\author{Yuki Oyama} 
\ead{oyama@shibaura-it.ac.jp}

\address{Department of Civil Engineering, Shibaura Institute of Technology, Tokyo, Japan}

\begin{abstract}
\small
Although the recursive logit (RL) model has been recently popular and has led to many applications and extensions, an important numerical issue with respect to the computation of value functions remains unsolved. 
This issue is particularly significant for model estimation, during which the parameters are updated every iteration and may violate the feasibility condition of the value function. To solve this numerical issue of the value function in the model estimation, this study performs an extensive analysis of a prism-constrained RL (Prism-RL) model proposed by \cite{oyama2019prism}, which has a path set constrained by the prism defined based upon a state-extended network representation. 
The numerical experiments have shown two important properties of the Prism-RL model for parameter estimation. First, the prism-based approach enables estimation regardless of the initial and true parameter values, even in cases where the original RL model cannot be estimated due to the numerical problem. We also successfully captured a positive effect of the presence of street green on pedestrian route choice in a real application. 
Second, the Prism-RL model achieved better fit and prediction performance than the RL model, by implicitly restricting paths with large detour or many loops. Defining the prism-based path set in a data-oriented manner, we demonstrated the possibility of the Prism-RL model describing more realistic route choice behavior. The capture of positive network attributes while retaining the diversity of path alternatives is important in many applications such as pedestrian route choice and sequential destination choice behavior, and thus the prism-based approach significantly extends the practical applicability of the RL model.
\end{abstract} 

\begin{keyword}
\small
Route choice analysis \sep recursive logit \sep prism constraint \sep pedestrian \sep GPS \sep city center
\end{keyword}

\end{frontmatter}

\newtheorem{thm}{Theorem}
\newtheorem{prop}{Proposition}
\newtheorem{defi}{Definition}


\section{Introduction}\noindent
Recursive logit (RL) models, which are derived based upon the Markov decision process, provide a computationally efficient way of modeling route choice behavior and can be consistently estimated \citep{Fosgerau2013RL, Mai2015NRL}. 
However, the evaluation of value functions required in the computation of such models remains challenging. The value function may not have a solution when the network contains cycles, depending on the utility value of the elemental link/node choice \citep[e.g., shown in the illustrative examples of][]{Oyama2017GRL, oyama2019prism}. This numerical problem is in particular significant in model estimation. During the estimation process, the model parameters that decide the utility values are updated with every iteration. Even if the initial and true parameters are feasible in terms of having a value function solution, parameters in the middle of the estimation may violate the feasibility condition of the value function. To somehow avoid this numerical issue, studies in the literature often included only negative network attributes (e.g., travel time or distance) and added a fixed penalty term for u-turns in the utility function \citep[e.g.,][]{Fosgerau2013RL, Mai2015NRL}. However, these ad hoc manipulations imply a limitation of the applicability of RL models. 
Recursive modeling of behavior is a general framework and its application is not limited to just route choice behavior, but also to various types of sequential decision making, such as destination sequences \citep{gao2021estimation} or mode chains \citep{de2019modelling}. 
The possible impacts of the urban/transport planning projects to be evaluated are not limited to negative ones. In recent urban design projects, for example, planners make better places to improve the walkability of city centers \citep[e.g.,][]{mehta2008walkable, mueller2020changing}, and the RL model can be a useful tool to evaluate possible positive impacts on pedestrian route choice whose path alternatives may be diverse. 
Therefore, to increase the practical applicability of RL models, it is important to solve their numerical issue and enable the capture of positive network attributes in utility.

As a key to solving the numerical issue of the evaluation of value functions for RL models, in the field of traffic assignment, \cite{oyama2019prism} recently proposed a prism-based path set restriction method. The method defines a state-extended network based on choice stages and restricts the set of states by introducing a prism constraint. Based on the restricted network, they redefined the value functions and proposed a prism-constrained RL (Prism-RL) model. Applying this model to Markovian traffic assignment (MTA), they showed that the incorporation of prism constraint makes the value function solvable regardless of the network conditions without the loss of efficiency (i.e., the implicit path enumeration) of the RL model, and also prevents the model from assigning excessive flows on cycles.
As such, the prism-based approach has the potential to solve the numerical problem of RL models; however, its applicability to model estimation has not yet been validated. The numerical issue of the value function is more significant in model estimation, and it is particularly difficult to successfully estimate RL models without numerical problems when we want to deal with many parameters or capture positive attributes. Therefore, it is important to validate the prism-based approach for the estimation of RL models. 

Furthermore, the key parameter of the prism-based approach is the so-called choice stage constraint, which defines the maximum number of choices a traveler can experience on the network. Although \cite{oyama2019prism} analyzed its impact on prediction performance, they did not detail how to define it in real applications. Moreover, since the choice stage constraint shapes the path set of the model, it may also affect the model estimation results, and thus the consistency property of the estimator may not be retained for the Prism-RL model.
Toward the practical application of the prism-based approach, the determination of this parameter and its trade-off between benefits and limitations need to be examined in more detail using real observations.

The objective of this study is to validate the applicability of the prism-based approach in the estimation of RL models. Starting with reviews of the original RL model and the Prism-RL model, we discuss in detail the numerical problem of the value function and how the prism-based approach solves it. We also add behavioral interpretation to the Prism-RL model. Then, we present a set of detailed numerical experiments to examine the properties of the Prism-RL model in comparison with the original RL models. In an experiment with simulated observations, we show that the Prism-RL model can be estimated and the true parameters are reproduced regardless of the starting points, while the RL model experiences the numerical issue of the value function during the estimation process. In addition, we present a real application to an urban pedestrian network, where pedestrians may take diverse paths and perceive not only negative attributes (i.e. travel costs) but also positive attractiveness of the streets. It is shown that the Prism-RL model can be actually estimated with a positive network attribute, while the RL model suffers from the numerical issue. We also compare the estimation results of the models including the nested version of the RL and Prism-RL models. Moreover, for the real application, we show a way to determine the choice stage constraint parameter based on real observations and examine its impact on the estimation, where limitations of the prism-based approach are also discussed. The contributions of this paper lie mainly in these extensive numerical experiments on the application of the prism-based approach to the estimation of RL models.

The structure of this paper is as follows. Section \ref{sec:RL} reviews the RL model and its recent advances, as well as its numerical problem. Then, in Section \ref{sec:method}, we describe the Prism-RL model proposed by \cite{oyama2019prism}, where we add a theoretical property and a behavioral interpretation to its value functions. Section \ref{sec:results} provides two types of numerical results: one uses simulated observations, and the other is a real application to pedestrian route choice analysis. Based on the numerical results, Section \ref{sec:direction} discusses two directions for further applications of the approach. Finally, we conclude the study in Section \ref{sec:conclusion}. The appendices also provide characteristics of the network and behavioral data used in the analysis and detailed computational reports.

\section{Numerical Issue of Recursive Logit Model}\label{sec:RL}\noindent 
This section briefly reviews the RL model \citep{Fosgerau2013RL} and its recent advances, and then discusses its numerical issue.

\subsection{Recursive Logit Model}
Consider a connected directed graph $G = (N, L)$ that is not assumed acyclic, where $N$ is the set of nodes and $L$ is the set of links. The RL model describes route choice behavior by a sequence of elemental state/action choices, based on a deterministic Markov decision process with i.i.d. Gumbel distributed rewards \citep{ziebart2008maximum, Rust1987}. 
Depending on the state definition, an RL model can be either node-based or link-based.
A node-based RL model provides a simpler description of route choice behavior with efficient computation and is often implemented in traffic assignment models \citep[e.g.,][]{Akamatsu1996MCA, oyama2022markovian}. In contrast, a link-based model describes a transition between two links, i.e. the geometric relationships of three nodes, and may capture more flexible mechanisms of behavior. For this reason, studies in the context of discrete choice analysis often use a link-based RL model \citep[e.g.,][]{Fosgerau2013RL, Mai2015NRL}, and we also used a link-based model in the numerical experiments in Section \ref{sec:results}.

A traveler in state $k$ is assumed to choose the next state (i.e., action) $a$ that maximizes the sum of instantaneous utility $u(a|k)$ and the expected downstream utility $V^d(a)$ to the destination (absorbing state) $d$. The utility $u(a|k)$ is further decomposed into the deterministic component $v(a|k)$ and the error component $\epsilon(a|k)$. The expected utility $V^d(k)$ is the value function of state $k$ that is formulated via Bellman equation:
\begin{equation}
    \label{eq:bellman}
    V^d(k) \equiv \mathbb{E}\left[\max_{a \in A(k)} \{v(a|k) + V^d(a) + \mu \epsilon(a|k) \} \right],
\end{equation}
where $A(k)$ is the set of states connected to (i.e., available actions for) state $k$, and $\mu$ is the scale of $\epsilon(a|k)$. With the distributional assumption $\epsilon(a|k)  \stackrel{iid}{\sim} {\rm Gumbel}(0, \mu)$ and by taking exponentials, (\ref{eq:bellman}) reduces to
\begin{equation}
    \label{eq:logsum}
    e^{\frac{1}{\mu} V^d(k)} = \sum_{a \in A(k)} e^{\frac{1}{\mu} \{v(a|k) + V^d(a)\}},
\end{equation}
which is a system of linear equations. In matrix form, (\ref{eq:logsum}) is compactly written as
\begin{equation}
    \label{eq:linearsystem}
    \mathbold{z}^d = \mathbf{M}\mathbold{z}^d + \mathbold{b}^d \Leftrightarrow \mathbold{z}^d = (\mathbf{I} - \mathbf{M})^{-1} \mathbold{b}^d
\end{equation}
where $z^d_k = e^{\frac{1}{\mu} V^d(k)}$, $M_{ka} = \delta(a|k) e^{\frac{1}{\mu} v(k|a)}$, and $\delta(a|k)$ is the state-action incidence. Also, $b^d_k$ equals one if $k = d$ and zero otherwise.
Finally, the RL model describes the choice probability of a path (i.e., state sequence) $\sigma = [k_1, \ldots, k_J]$ by a product of elemental choice probabilities:
\begin{equation}
    \label{eq:choiceprob}
    P(\sigma) = \prod_{j=1}^{J-1} p^d(k_{j+1}|k_j) = \prod_{j=1}^{J-1} \frac{
    e^{\frac{1}{\mu} \{v(k_{j+1}|k_j) + V^d(k_{j+1})\}}
    }{
    \sum_{a \in A(k_j)} e^{\frac{1}{\mu} \{v(a|k_j) + V^d(a)\}}
    }.
\end{equation}
Note that the equivalence of (\ref{eq:choiceprob}) with the multinomial logit (MNL) type route choice model with the unrestricted path set has been proved \citep{Akamatsu1996MCA, Fosgerau2013RL}. We refer the reader to \cite{Fosgerau2013RL} for more details of the RL model.

Implicitly considering the unrestricted path set, the RL model is computationally efficient and can be consistently estimated, and has been recently popular and extended to a variety of variants, such as the nested RL (NRL) model \citep{Mai2015NRL}, the network multivariate extreme value (NMEV) model \citep{Mai2016NGEV}, a mixture model \citep{Mai2018MRL}, incorporation of a spatial discount factor \citep{Oyama2017GRL}, of travel information \citep{de2020route}, and of network uncertainty \citep{mai2021route}.
\cite{Oyama2018ped} and \cite{Oijen2020wifi} proposed network-free parameter estimation methods which are based on raw trajectory data of GPS locations and WiFi traces, respectively. The RL model is also the fundamental model of the MTA \citep{Akamatsu1996MCA, Akamatsu1997Entropy, Baillon2008MCA}. \cite{oyama2022markovian} recently presented a path algebra for MTA that unifies the logit- and NMEV-based RL models, as well as the deterministic shortest path assignment, and also performed a theoretical analysis of the Markovian traffic equilibrium based on the NMEV-RL model. Furthermore, applications of MTA in the machine learning field have been recently studied  \citep{Saerens2009randomized, kivimaki2020maximum}.

\subsection{Numerical Issue on Value Function}\label{sec:RL_value}
Despite its nice properties and many applications, the RL model still has a fundamental numerical issue to be addressed: the existence of its solution depends on the structure of the network, the magnitude of utility, and the values of the model parameters. When the network contains cycles, the value function may not be solved depending on the model parameters. Mathematically, $(\mathbf{I} - \mathbf{M})^{-1}$ exists if the spectral radius $\rho(\mathbf{M})$, i.e., the maximum absolute of the eigenvalue, of the matrix $\mathbf{M}$ is strictly smaller than one \citep[e.g.,][]{varga1962iterative, atkinson2008introduction}. 
\cite{mai2022undiscounted} recently showed that under the condition of $\sum_{a \in A(k)} M_{ka} < 1, \forall k$, the Bellman equation of the RL model is a contraction mapping, and thus the linear system (\ref{eq:linearsystem}) has a unique solution and can be efficiently computed. In their study, the theoretical result for the NRL model was also derived.

This numerical problem of the value function is in particular significant during model estimation.
The estimation of RL models requires a structural estimation method, and the nested fixed point (NFXP) algorithm of \cite{Rust1987} has often been used. The structural estimation involves two problems: an outer nonlinear optimization to maximize the likelihood function over the parameter space and an inner problem to solve the value function for each parameter. Whether the value function can be solved depends on the magnitude of utility, hence the parameter value. That is, to successfully estimate an RL model, all parameter values searched by an outer nonlinear optimization algorithm have to satisfy the feasibility condition for the value function.

To somehow satisfy the conditions during the estimation, previous studies on RL models often included in utility only negative network attributes and/or add a fixed large penalty for u-turns \citep[e.g.,][]{Fosgerau2013RL, Mai2015NRL}. \cite{kaneko2018route} incorporated the probability of link awareness to reduce the utilities of links with small observed traffic.
However, there is no guarantee that such ad hoc manipulations of the utility function always resolve the numerical issue during the estimation. \cite{Oyama2017GRL} (in Section 4.3) and \cite{oyama2019prism} (in Appendix B) analyzed the spectral radius condition and showed that the value function may not be solved even when only negative attributes are contained in the utility function. There are some studies that successfully estimated RL models with positive attributes \citep[e.g.,][]{zimmermann2017bike, gao2021estimation}, but it is not ensured that the feasibility condition of the value function is always satisfied during the parameter search. It is also difficult for RL models to find a good initial value for a nonlinear optimization algorithm that would free us from the numerical issue during the estimation. Therefore, an estimation method for the RL models that does not depend on network conditions or initial values is still needed to increase their applicability.

Note that it is known that the Bellman equation is a contraction mapping when we have a discount factor smaller than one. \cite{Oyama2017GRL} incorporated the discount factor into the RL model and analyzed the myopic route choice behavior of drivers in an extremely congested network during a disaster. However, a large discount on the value function seems to be unlikely to occur in ordinary networks on a real scale, and studies on route choice modeling and traffic assignment generally deal with an undiscounted case \citep[e.g.,][]{Akamatsu1996MCA, Fosgerau2013RL, mai2022undiscounted}. Therefore, although the prism-based approach may also be suitable with the discounted model, this study focuses on the undiscounted case (i.e., the discount factor equals one) in which the numerical issue related to the value function is important.

\section{Prism-Constrained Recursive Logit Model}\label{sec:method}\noindent
This section describes the Prism-RL model proposed by \cite{oyama2019prism}, to which we add a behavioral interpretation and a theoretical property of its value functions, as well as model estimation.

\subsection{Prism-based Path Set Restriction}\label{sec:prism}
A state-extended network is constructed based on the spatial network $G$, where a state $s$ is defined based on a pair of \textit{choice stage} $t$ and \textit{place}\footnote{We use the term \textit{place} for generalization of node and link in the model description.} (a node or link) $k$, that is, $s = (t, k)$. A choice stage is the timing of a traveler's decision making and does not indicate a time. It is assumed that the instantaneous utility $v(a|k)$ does not vary by choice stage, which means that the Prism-RL model is time-independent\footnote{This study focuses on a time-independent model for the comparison with the original RL model, while the extension to a time-space network description is straightforward \citep{Oyama2016HKSTS}}.
Based on the state-extended network, we further introduce a choice stage constraint $T$ that is the maximum number of choices a traveler can experience on the network, whose interpretation differs from the definition of the original network \citep[see Table 1 of][]{oyama2019prism}. The constraint $T$ can be a single scalar, or defined by destination ($T = \{T_{d}\}$) or by origin-destination (OD) pair ($T = \{T_{od}\}$).
Under the constraint, a traveler who wants to travel to the destination $d$ must arrive there at or before the choice stage $T$, that is, her terminal state is always $s_{T} = (T, d)$. We use this fact to reduce the size of states in the network and restrict the set of feasible paths.

Let $D^d(k)$ denote the minimum number of steps (i.e., choice stages) to take from $k$ to $d$, which can be computed with a shortest path algorithm on $G$. The existence condition $I^d(s)$ of state $s = (t,k)$ is then defined as
\begin{equation}
    I^d(t,k) = 
    \left\{
    \begin{array}{c c}
    1,\:& {\rm if} \:\: D^d(k) \leq T-t\\
    0 ,\:& {\rm otherwise}.\\
    \end{array}
    \right.
    \hspace{\fill}
    \forall t \in \{0, \ldots, T-1\}
    \hspace{0.5cm}
\label{eq:I}
\end{equation}
which means that a traveler is allowed to be in place $k$ only at $t$ that satisfies $I^d(t,k) = 1$. The state set $S^d_t$ at choice stage $t$ thus reduces to $S^d_t \equiv \{s = (t,k) | I^d(s) = 1 \}$. Moreover, a state transition is possible only when two states exist and are spatially connected, i.e., the state connection condition $\Delta^d(s'|s) = \Delta^d_t(a|k)$ between states $s = (t, k)$ and $s' = (t+1, a)$ is
\begin{equation}
    \label{eq:Delta}
    \Delta^d_t(a|k) = I^d(t,k) \delta(a|k) I^d(t+1, a).
\end{equation}
The set of edges (connected pairs of states) $E^d_t$ is also restricted to $E^d_t \equiv \{(s, s') = ((t, k), (t+1, a)) | \Delta^d(s'|s) = 1 \}$. 
Constraint $\Delta$ implies that the set of places available to a traveler varies by choice stage $t$ even if she is in the same place. Thus, feasible paths must include only states in the reduced network. The set of paths that a traveler can choose results in the formation of a prism \citep{Hagerstrand1970}, which exhibits a sphere of possible behavior of a traveler in the state-extended network. 

Two remarks are in order here. First, the choice stage constraint $T$ is a hyperparameter, and the information from real observations, such as the detour characteristics or the maximum number of choice stages in a path, can be utilized for the decision, for which we later show an example in the real case study in Section \ref{sec:application}. Depending on the prior information, $T$ can be defined for each destination or OD pair, or as a single value common among them.
Second, while we have discussed above the prism that has only a vertex of the final state $(T, d)$, it can be defined based also upon the initial state $s_0 = (0, o)$ where $o$ is the origin. In this case, an additional constraint is required for the existence of the state, and (\ref{eq:I}) is replaced by
\begin{equation}
    I^{od}(t,k) = 
    \left\{
    \begin{array}{c c}
    1,\:& {\rm if} \:\: D^d(k) \leq T-t, D^o(k) \leq t\\
    0 ,\:& {\rm otherwise}.\\
    \end{array}
    \right.
    \hspace{\fill}
    \forall t \in \{0, \ldots, T-1\}
    \hspace{0.5cm}
\label{eq:Irev}
\end{equation}
where $D^o(k)$ denotes the minimum number of steps to take from $o$ to $k$. This doubly constrained prism approach describes a more precise behavioral sphere and can reduce the network states to consider, but the procedure has to be repeated as many times as the number of OD pairs, which may be computationally expensive in a large-scale network. 


\subsection{Prism-Constrained RL Model}\noindent
We then introduce the Prism-RL model, which is an RL model whose path set is restricted by the prism constraint of Section \ref{sec:prism}. The main difference from the original RL model is that the Prism-RL model defines the value function $V^d(s) = V^d(t, k)$ for each state $s \in S^d \equiv \{S^d_0, \ldots, S^d_{T}\}$ in the extended network.

As a result, (\ref{eq:logsum}) and (\ref{eq:linearsystem}) of the RL model are replaced by
\begin{equation}
    \label{eq:logsum-p}
    e^{\frac{1}{\mu} V^d(t,k)} = \sum_{a \in A(k)} \Delta^d_t(a|k) e^{\frac{1}{\mu} \{v(a|k) + V^d(t+1, a)\}}
\end{equation}
and
\begin{equation}
    \label{eq:linearsystem-p}
    \mathbold{z}^d_t = \mathbf{M}^{'d}_{t} \mathbold{z}^d_{t+1} + \mathbold{b}^d 
    ~~~\forall t \in \{0, \ldots T-1\},
\end{equation}
where $z^d_{t,k} = e^{\frac{1}{\mu} V^d(t,k)}$, and $M^{'d}_{t,ka} = \Delta^d_t(a|k) e^{\frac{1}{\mu} v(a|k)}$. 

\begin{figure}[tbh]
	\begin{center}
		\includegraphics[width=15cm]{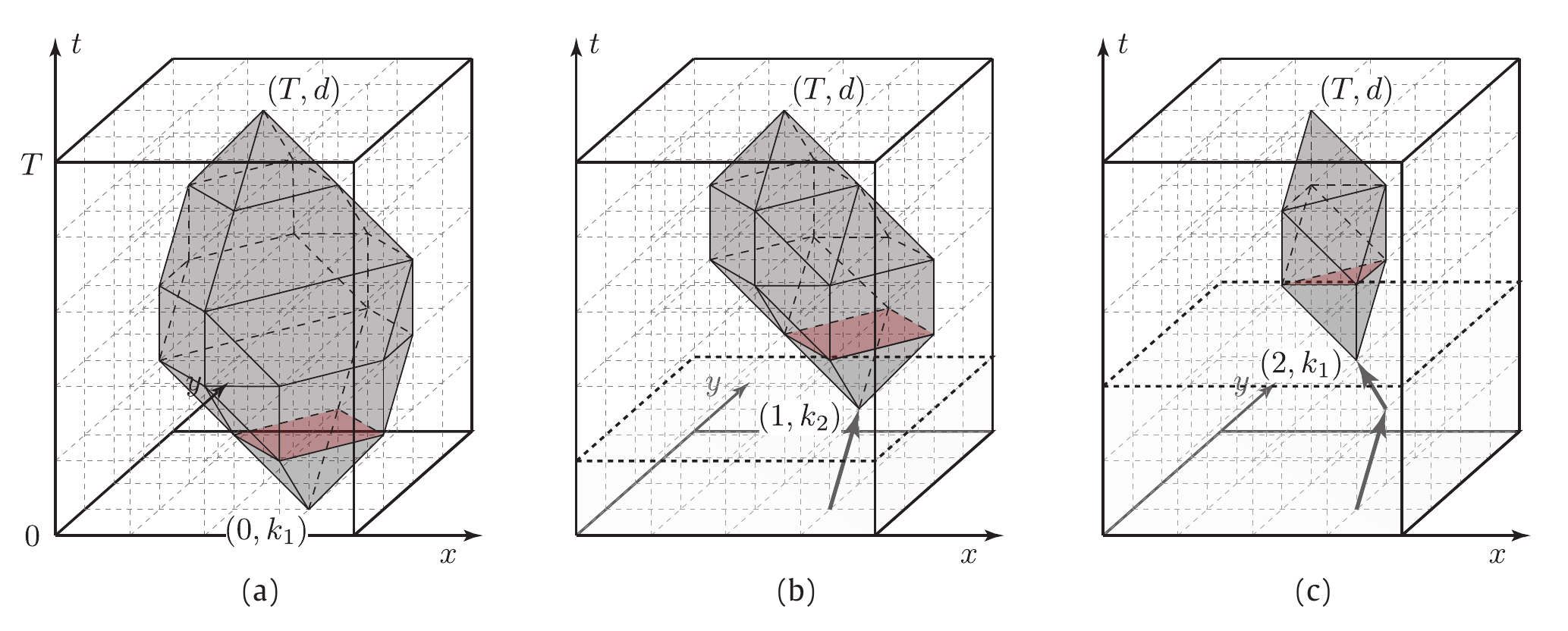}
		\caption{Prism representing the restricted set of states conditional on each of current states (a) $(0, k_1)$, (b) $(1, k_2)$, and (c) $(2, k_1)$, where travelers choose routes on a $5 \times 5$ grid network and the choice stage constraint is $T=5$. 
		The gray prism is a collection of future movements available to travelers in each state, and the red plane indicates the possible action space (i.e., the choice set faced in the current state). The arrows indicate the actions taken at the previous choice stages.
		The value function $V^d(s)$ evaluates all the feasible paths within the prism and thus takes different values by choice stages even in the same place; in this example, the panels (a) and (c) illustrate that $V^d(0, k_1) \neq V^d(2, k_1)$. 
		Note that this figure shows a doubly-constrained case.}
		\label{fig:prism} 
	\end{center}
\end{figure}

To give a behavioral interpretation of the value function, in Figure \ref{fig:prism} we illustrate how a prism is defined for different states, for a route choice example on a $5 \times 5$ grid network \citep[the same network as that of Fig.1 in][]{oyama2019prism}. The traveler in each state is faced with a choice for next action, and the action space (i.e., the choice set) is a restricted plane by a prism. A prism is a collection of all future movements available to travelers in each state, and the value function $V^d(t, k)$ evaluates the prism defined based on the current state $(t,k)$ and the final state $(T,d)$ by the expected maximum utility. Therefore, the prism's form and the value function may be different by choice stages even in the same place: $V^d(t, k) \neq V^d(t', k)$, $t' > t$, and travelers at the upper choice stage $t'$ have a higher probability of taking actions that lead them to the destination more efficiently.  

\begin{figure}[tbh]
	\begin{center}
		\includegraphics[width=13cm]{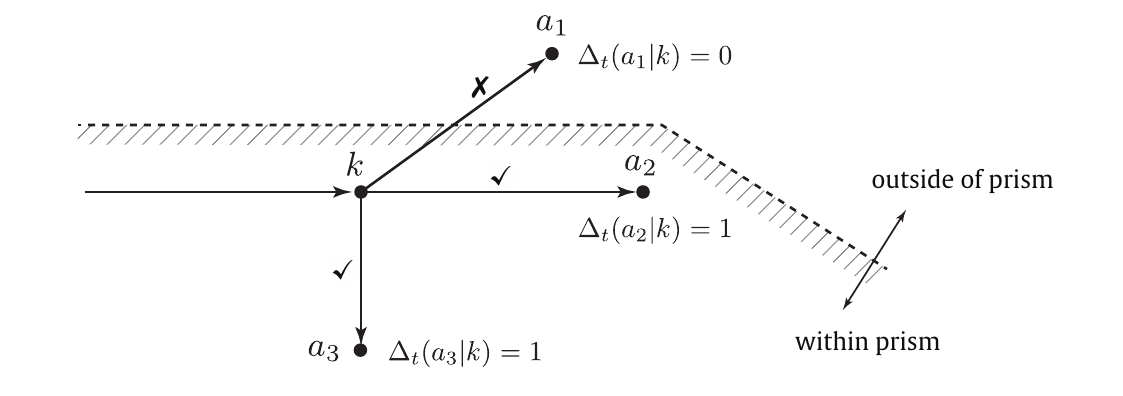}
		\caption{Route choice behavior constrained by the prism (reproduction of Fig.2 of \cite{oyama2019prism} with minor edits): In this example, node $k$ is connected to three nodes $A(k) = \{a_1, a_2, a_3\}$, but state $(t+1, a_1)$ is outside of the prism, i.e., $I(t+1, a_1) = 0$. At node $k$ and choice stage $t$, therefore, the choice of $a_1$ is violated ($\Delta_t(a_1|k) = 0$), and only the transitions to $a_2$ or $a_3$ are allowed ($\Delta_t(a_2|k) = \Delta_t(a_3|k) = 1$).}
		\label{fig:Delta} 
	\end{center}
\end{figure}

Finally, with the value functions defined by choice stages, the Prism-RL model describes the choice probability $p^d(s'|s) = p^d_t(a|k)$ of state $s' = (t+1, a)$ given a current state $s = (t,k)$ as
\begin{equation}
    \label{eq:prism_prob}
    p^d_t(a|k) = \frac{
    \Delta^d_t(a|k) e^{\frac{1}{\mu} \{v(a|k) + V^d(t+1, a)\}}
    }{
    \sum_{a' \in A(k)} \Delta^d_t(a'|k) e^{\frac{1}{\mu} \{v(a'|k) + V^d(t+1, a')\}}
    }.
\end{equation}
where we have the prism constraint $\Delta^d_t(a|k)$, meaning that no state transition going outside the prism is allowed even if the two states are spatially connected (Figure \ref{fig:Delta}). 

\subsection{Value Function Computation}\label{sec:prism_value}
Thanks to the prism constraint and the state-specific definition of the value function, we no longer need to solve the linear system (\ref{eq:linearsystem}), and a unique solution always exists regardless of the network structure or utility specification. To further discuss this property, we recursively substitute (\ref{eq:linearsystem-p}) into its right-hand side and have (we omit $d$ in the equations of this subsection for simplicity as they are common for all $d \in \mathcal{D}$):
\begin{align}
    \label{eq:linearsystem-expansion}
    \mathbold{z}_t &= \mathbf{M}^{'}_{t} \mathbold{z}_{t+1} + \mathbold{b} \nonumber\\
    &= \mathbf{M}^{'}_{t} (\mathbf{M}^{'}_{t+1} \mathbold{z}_{t+2} + \mathbold{b}) + \mathbold{b} \nonumber\\
    &= \mathbf{M}^{'}_{t}\mathbf{M}^{'}_{t+1} (\mathbf{M}^{'}_{t+2} \mathbold{z}_{t+3} + \mathbold{b}) + (\mathbf{I} + \mathbf{M}^{'}_{t}) \mathbold{b} \nonumber\\
    &= \cdots \nonumber\\
    &= (\mathbf{I} + \sum_{r=t}^{T-1} \prod_{s=t}^{r} \mathbf{M}^{'}_{s}) \mathbold{b}.
\end{align}
where $\mathbold{z}_{T} = \mathbold{b}$.
Since $M^{'}_{t,ka} \le M_{ka}, \forall t \in \{0,\ldots,T-1\}$ by the definition of $\Delta$ (\ref{eq:Delta}), there always exists a scalar $C \in \mathbb{R}_{\ge 0}$ such that 
\begin{align}
    \label{eq:Mineq}
    \prod_{s=t}^{r} \mathbf{M}^{'}_{s} \le \mathbf{M}^{r-t+1} \le C,
    ~~~\forall 0 \le t \le r \le T-1,
\end{align}
in other words, as long as $T = \max(r-t+1)$ is a finite value, $\prod_{s=t}^{r} \mathbf{M}^{'}_{s}$ is upper-bounded by a real number $C$. Consequently, $\mathbold{z}_t$ is also a real vector for all $t \in \{0, \ldots, T\}$.

To solve (\ref{eq:linearsystem-p}), we can apply a simple backward calculation with $T$ iterations. We initialize the vector of the value function: $z_{T,d} = 1$ and $z_{t,k} = 0, \forall (t,k) \neq (T,d)$. 
We start at $t = T-1$ and update the value function by
\begin{equation}
    \label{eq:linearsystem-p-alg}
    \mathbold{z}_t \leftarrow \mathbf{M}^{'}_{t} \mathbold{z}_{t+1} + \mathbold{b},
\end{equation}
and repeat updating backward in choice stage until $t = 0$, i.e., $\mathbold{z}_0$ is computed. 



It is worth noting that this solution method does not depend on the linearity of the model. 
It is also applicable to nonlinear models, such as the NRL model \citep{Mai2015NRL} and the NMEV model \citep{Mai2016NGEV, oyama2022markovian}. In the case of the prism-constrained NRL (Prism-NRL) model, for instance, (\ref{eq:linearsystem-p}) is replaced by
\begin{equation}
    \label{eq:nonlinearsystem-p}
    \mathbold{z}^{\rm NRL}_t = [\mathbf{M}^{'}_{t} \circ \mathbf{X} (\mathbold{z}^{\rm NRL}_{t+1})]\mathbold{e} + \mathbold{b} 
    ~~~\forall t \in \{0, \ldots T-1\},
\end{equation}
where $X(\mathbold{z}^{\rm NRL}_{t})_{ka} = (z^{\rm NRL}_{t,a})^{\mu_a/\mu_k}$, and $\mu_{k}$ is the place-specific scale parameter to capture the correlation among utilities. The vector $\mathbold{e}$ is an identity vector, and $\circ$ is the element-by-element product. 
Therefore, (\ref{eq:nonlinearsystem-p}) can be solved in the same way as the Prism-RL model, i.e., by the iterate update of
\begin{equation}
    \label{eq:nonlinearsystem-p-alg}
    \mathbold{z}^{\rm NRL}_t \leftarrow [\mathbf{M}^{'}_{t} \circ \mathbf{X} (\mathbold{z}^{\rm NRL}_{t+1})]\mathbold{e} + \mathbold{b}
\end{equation}
from $t=T-1$ to $t=0$. It can be naturally stated from the discussion of Prism-RL model that the value function $\mathbold{z}^{\rm NRL}_t, ~ \forall t \in \{0, \ldots T\}$ for the Prism-NRL model is also upper-bounded by a real vector, which allows for a computation of the value function regardless of network or parameter conditions.

\subsection{Path Translation and Model Estimation}\label{sec:estimation}
The parameters of the Prism-RL model are estimated by the maximum likelihood estimation. Assume that we have route choice observations $\sigma_n = [k_0, \ldots, k_{J_n}]$, $n \in \{1, \ldots, N\}$, where an observed path $\sigma_n$ is a sequence of places (nodes or links) of length $J_n$, and the last element $k_{J_n} = d_n$ corresponds to its destination.
To translate the observations into those for the Prism-RL model, we first set the choice state constraint $T$ to a value equal to or greater than $\bar{J} \equiv \max_n J_n$ so that all observed paths are contained in the prism.
Then $\sigma_n$ is translated into the sequence of pairs of choice stage and place, i.e., $\sigma^*_n = [(0, k_0), \ldots, (J_n, d_n), (J_n+1, d_n), \ldots (T, d_n)]$, where the traveler is assumed to be at the destination $d_n$ until $T$ after arriving at $d_n$, thus $k_{J_n} = \cdots = k_{T} = d_n$.
The transition probability from states $(t, d_n)$ to $(t+1, d_n)$ is fixed to one.
Note that this translation does not change the nature of the data and its behavioral interpretation. The log-likelihood function of the Prism-RL model is
\begin{eqnarray}
    \label{eq:like}
    LL(\beta; \sigma^*) &\equiv&
    \log \prod^N_{n=1} P(\sigma_n) \nonumber\\
    &=& 
    \sum^N_{n=1} \sum^{T-1}_{t=0} \log p^{d_n}_t(k_{t+1}|k_t) \nonumber\\
    &=& \frac{1}{\mu} \sum^N_{n=1} \sum^{T-1}_{t=0} \left[
    v(k_{t+1}|k_t) + V^{d_n}(t+1, k_{t+1}) - V^{d_n}(t, k_t)
    \right]
\end{eqnarray}
where $\beta$ is the vector of model parameters, and $v(a|k) = v(\mathbold{x}_{a|k}, \beta)$ is a function of $\beta$ and a vector of observed attributes $\mathbold{x}_{a|k}$ of place pair $(k, a)$. 
The maximization problem of (\ref{eq:like}) can be solved by the same optimization algorithms for the original RL models \citep[e.g.,][]{Fosgerau2013RL, Mai2015NRL}. This study applied the NFXP algorithm \citep{Rust1987}, in which the evaluation of the value function is performed every time the parameter is updated. The parameter search is performed using the BFGS method, which is a quasi-Newton-type nonlinear optimization algorithm.

\section{Numerical Results}\label{sec:results}\noindent
In this section, we first present an experiment using observations simulated in the Sioux Falls network. In the experiment, we examine the parameter reproducibility and the estimation processes of the RL and Prism-RL models. We then provide a real application result in the case study of pedestrian route choice, where we show that the prism-based approach allows us to capture a positive attribute of green presence on the streets. We also examine the determination of choice stage constraint $T$ and the trade-off between the benefits and limitations of the prism-based approach. 

We used the link-based formulation of the models and defined a prism for each destination. 
All models have been implemented in Python 3.6 on a machine with 14 core Intel Xeon W processors (2.5 GHz) and 64 GB of RAM. We implemented modeling, estimation, and validation by writing our own Python code, where for parameter search, we used the BFGS algorithm of the SciPy \textit{optimize} module.

\subsection{Experiment with Simulated Observations}\label{sec:siouxfalls}
First, we report several results using simulated observations to show that the proposed Prism-RL model actually solves the numerical problem of the RL model. In the experiment, we use the Sioux-Falls network \citep{SiouxFalls2016} and define the utility function as
\begin{equation}
    \label{eq:util}
    v(a|k) = (\beta_{\rm{len}} + \beta_{\rm{cap}} {\rm Capacity}_a){\rm Length}_a - 10 {\rm Uturn}_{a|k}
\end{equation}
where ${\rm Length}_a$ is the length of link $a$, and ${\rm Capacity}_a$ is its capacity divided by the maximum link capacity in the network, which are available from the dataset (see Figure \ref{fig:net} in \ref{app:network})\footnote{Note that we use the variables of length and capacity just for the sake of convenience. We used only simulated observations for the Sioux Falls network experiment, and the behavioral interpretations of the parameters are not of interest.}. The effect of ${\rm Capacity}_a$ is captured by an interaction term with ${\rm Length}_a$ so that the utilities are link-additive. 
We also add a fixed negative u-turn penalty ${\rm Uturn}_{a|k}$, which follows previous studies of RL models \citep[e.g.,][]{Fosgerau2013RL}. Here, as the main difference from the literature, we consider a possible positive effect of capacity on utility, i.e., $\beta_{\rm cap}$ can be greater than zero, while length has a negative effect $\beta_{\rm len} < 0$. Previous RL models have not been able to include positive utilities due to the numerical issue with the evaluation of the value function.

In the experiment, we always use observations simulated by the RL model with a known vector of true parameters $\tilde{\beta}$ and compare the estimation of the RL and Prism-RL models. We add the dimension of the choice stage to the observations for the estimation of the Prism-RL model, by the path translation we explained in Section \ref{sec:estimation}, but the nature of the data does not change. We set the choice stage constraint $T = 15$\footnote{We additionally tested the cases with $T = 10, 25, 50, 75, 100$. The value of $T$ did not affect the estimation result in the Sioux Falls experiment, while the computational time increased approximately linearly with the increase in $T$. See \ref{app:SF_difT} for details.}, which is common for all destinations. By implementing a Monte Carlo simulation, we generate 1,000 observations for each origin-destination (OD) pair. We considered four destinations and six origins, i.e., 24 OD pairs in total. Note that we did not observe any path with a loop where a repeated link is considered a loop.

The summary of the experiments provided in this section is as follows.
\renewcommand{\theenumi}{\arabic{enumi}}
\renewcommand{\labelenumi}{(\theenumi)}
\begin{enumerate}
    \item We compare the estimation results of the RL and Prism-RL models. Two cases are tested: (a) $\beta_{\rm cap} < 0$ and (b) $\beta_{\rm cap} > 0$. The objective of this experiment is to validate the potential of the Prism-RL model reproducing the true parameter of the RL model, even in the case in which a positive utility effect is considered.
    \item We then investigate the estimation processes of the models by visualizing the parameter update histories. This experiment focuses on a case with $\beta_{\rm cap} > 0$. We see how the RL model faces with the numerical issue during the estimation and how it is addressed during the estimation of the Prism-RL model.
\end{enumerate}

\subsubsection{Parameter Reproducibility}\label{sec:reproducibility}
This experiment validates the parameter reproducibility of the Prism-RL model. We divided the observations into 10 samples, each of which thus had 2,400 observations. The RL model and the Prism-RL model are estimated for every sample. The initial parameter values for the estimation were set to $(\beta_{\rm len}, \beta_{\rm cap}) = (-1.0, -1.0)$.

First, we set the true values of the parameters in (\ref{eq:util}) to $(\tilde{\beta}_{\rm len}, \tilde{\beta}_{\rm cap}) = (-2.0, -1.5)$, both of which are negative. Table \ref{tb:estimate1} reports the estimation results. In this case, the two models returned the same results. All estimates and standard errors of the RL and Prism-RL models were exactly equal to four decimal places. As shown in Table \ref{tb:estimate1}, the true parameters are reproduced well and all estimates of the 10 samples are not significantly different from the true values at the 5\% significance level. These results suggest that the Prism-RL model is able to sufficiently reproduce the parameters of the RL model.
Note that the reason that the two models obtained the same results seems to be that the prism defined in this experiment sufficiently covered the distributional region of the RL model's probability.

\begin{table}[thb]
	\centering 
	\footnotesize
	\caption{Estimation results with the simulated observations. The true parameter values are $(\tilde{\beta}_{\rm len}, \tilde{\beta}_{\rm cap}) = (-2.0, -1.5)$. All estimates and standard errors of the RL and Prism-RL models were exactly equal to four decimal places.}
	\label{tb:estimate1}
	\begin{tabular*}{\hsize}{@{\extracolsep{\fill}}crrrrrr@{}}
		\toprule
		Sample & $\hat{\beta}_{\rm len}$ & std.err. & t-test & $\hat{\beta}_{\rm cap}$ & std.err. & t-test \\
		\midrule
        1 & -2.013 & 0.108 & 0.118 & -1.479 & 0.150 & -0.137 \\
        2 & -2.004 & 0.048 & 0.090 & -1.519 & 0.057 & 0.332 \\
        3 & -2.019 & 0.071 & 0.274 & -1.521 & 0.124 & 0.173 \\
        4 & -2.046 & 0.045 & 1.033 & -1.506 & 0.056 & 0.104 \\
        5 & -2.003 & 0.114 & 0.026 & -1.490 & 0.115 & -0.086 \\
        6 & -2.028 & 0.075 & 0.368 & -1.551 & 0.069 & 0.743 \\
        7 & -2.001 & 0.048 & 0.024 & -1.492 & 0.058 & -0.132 \\
        8 & -1.979 & 0.051 & -0.415 & -1.497 & 0.062 & -0.045 \\
        9 & -2.060 & 0.046 & 1.307 & -1.537 & 0.057 & 0.653 \\
        10 & -1.999 & 0.072 & -0.016 & -1.499 & 0.063 & -0.016 \\
        \midrule
        Average & -2.015 & 0.068 & 0.281 & -1.509 & 0.081 & 0.159 \\
		\bottomrule
	\end{tabular*}
\end{table}

Next, we set the true parameters to $(\tilde{\beta}_{\rm len}, \tilde{\beta}_{\rm cap}) = (-2.5, 2.0)$ to include a positive attribute. 
In this case, unlike the previous case where both parameters are negative, we observed a clear difference between the results of the two models. The RL model did not converge for all samples; parameters were updated to ones with which the value function was not solvable during the estimation process and failed to obtain the estimation results, even though the true value is a feasible solution of the RL model.
In contrast, we succeeded in estimating the Prism-RL model for all samples. Table \ref{tb:estimate2_Prism} reports the estimation result of the Prism-RL model. 
Like in the previous experiment, all estimates of the 10 samples are close to their true values, and they are not significantly different from their true values at the 5\% significance level. These results validate that the Prism-RL model solves the numerical problem of the RL model regarding the evaluation of the value function and captures a positive variable effect on utility.

\begin{table}[htb]
	\centering 
	\footnotesize
	\caption{Estimation results of the Prism-RL model with the simulated observations. The true parameter values are $(\tilde{\beta}_{\rm len}, \tilde{\beta}_{\rm cap}) = (-2.5, 2.0)$. We did not obtain results for the RL model for any sample.}
	\label{tb:estimate2_Prism}
	\begin{tabular*}{\hsize}{@{\extracolsep{\fill}}crrrrrr@{}}
		\toprule
		Sample & $\hat{\beta}_{\rm len}$ & std.err. & t-test & $\hat{\beta}_{\rm cap}$ & std.err. & t-test \\
		\midrule
        1 & -2.509 & 0.064 & 0.139 & 2.002 & 0.525 & -0.004 \\
        2 & -2.539 & 0.049 & 0.790 & 2.005 & 0.051 & -0.090 \\
        3 & -2.416 & 0.066 & -1.282 & 1.930 & 0.082 & 0.859 \\
        4 & -2.515 & 0.449 & 0.033 & 2.020 & 0.079 & -0.248 \\
        5 & -2.448 & 0.805 & -0.065 & 1.949 & 0.159 & 0.321 \\
        6 & -2.453 & 0.066 & -0.711 & 1.960 & 0.066 & 0.614 \\
        7 & -2.526 & 0.211 & 0.124 & 2.028 & 0.224 & -0.125 \\
        8 & -2.483 & 0.300 & -0.057 & 1.971 & 0.076 & 0.388 \\
        9 & -2.442 & 0.053 & -1.104 & 1.926 & 0.051 & 1.435 \\
        10 & -2.470 & 0.348 & -0.086 & 1.976 & 0.578 & 0.041 \\
        \midrule
        Average & -2.480 & 0.241 & -0.222 & 1.977 & 0.189 & 0.319 \\
		\bottomrule
	\end{tabular*}
\end{table}

\subsubsection{Estimation Process}\label{sec:stability}
\begin{figure}[h] 
	\begin{center}
		\includegraphics[width=14cm]{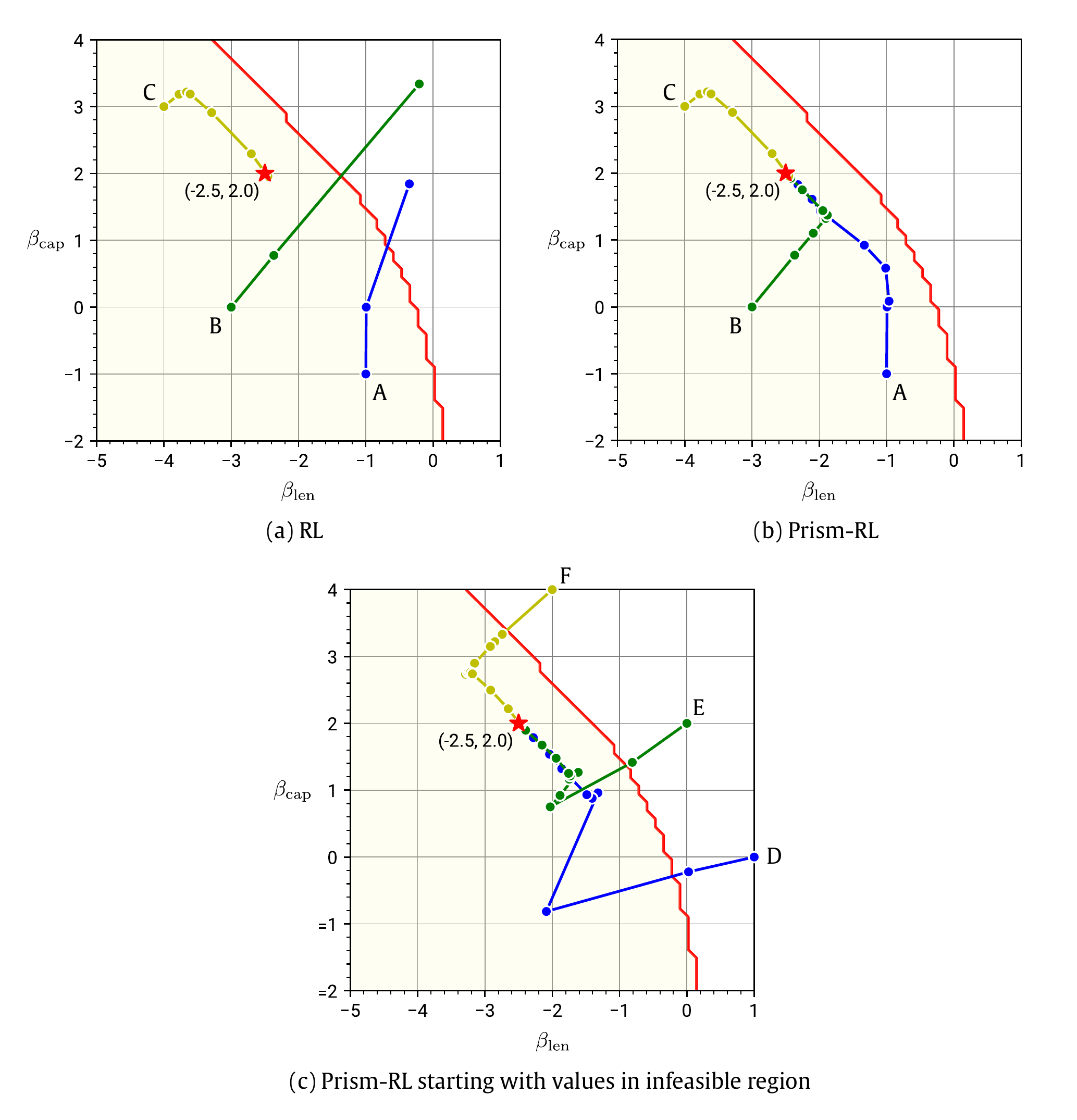}
		\caption{Parameter estimation processes of (a) RL model and (b) Prism-RL model tested with three different starting points A, B and C, and of (c) Prism-RL model with those in the infeasible region D, E and F. The yellowed area in the graph is the feasible region for the value function of the RL model.}
		\label{fig:feasibility} 
	\end{center}
\end{figure}

To understand how the numerical problem of the RL model occurs during the estimation, we visualized its estimation process with different initial values.
We used all observations at once for this experiment. Again, as in the previous experiment, we set the true values of the parameters in (\ref{eq:util}) to $(\tilde{\beta}_{\rm len}, \tilde{\beta}_{\rm cap}) = (-2.5, 2.0)$ including a positive attribute. 
With three different initial points A $(-1, -1)$, B $(-3, 0)$, and C $(-4, 3)$, we compared the estimation processes of the RL and Prism-RL models (Figure \ref{fig:feasibility}a,b). The \textit{feasibile region} is defined here as the parameter space where the value function of the RL model has a solution.
When the initial point was C $(-4, 3)$, for both the RL and Prism-RL models the nonlinear algorithm converged to the true solution. 
However, in cases A and B, the parameters were updated to the infeasible region of the RL model during the estimation process (blue and green trajectories in Figure \ref{fig:feasibility}a), at which the estimation failed. In contrast, the Prism-RL model did not experience the numerical issue. Regardless of the initial point, the Prism-RL model was estimated with the true value. During the estimation process, no fluctuation or update of the parameter to the RL infeasible region was not observed (Figure \ref{fig:feasibility}b).

Moreover, we tested the estimation of the Prism-RL model with a starting point set to a value outside of the RL feasible region. The result is shown in Figure \ref{fig:feasibility}(c). We tested three initial values D $(1, 0)$, E $(0, 2)$ and F $(-2, 4)$, with which the RL model could not be estimated due to the numerical problem of the value function. 
Figure \ref{fig:feasibility}(c) demonstrates that the Prism-RL model is estimable even with the initial values within the infeasible region for the value function of the RL model. This fact suggests that it is possible to estimate the Prism-RL model even if the parameter is updated outside of the RL feasible region during the estimation process.

\subsection{Pedestrian Route Choice Application}\label{sec:application}
To validate the usefulness of the prism-based approach in real applications, we present here a case study of pedestrian route choice. It is assumed that, unlike other modes of transportation, pedestrians have more freedom to choose their route and may be affected by positive attributes (i.e., attractiveness) of streets.
We use the GPS data collected through the Probe Person Survey, a complementary survey of the Sixth Tokyo Metropolitan Region Person Trip Survey \citep{TokyoPT}. 
This study focuses on the pedestrian network of a mile square centered on the Kannai Station, Yokohama city, Japan (Figure \ref{fig:kannai} in \ref{app:ppdata}). Yokohama is the second largest city in Japan by population. The Kannai district is popular for strolling; it is close to the sea and there are many places for refreshments, such as parks. The pedestrian network contains 724 nodes and 2398 links with 8434 link pairs. 

We analyze some characteristics of the observed paths in \ref{app:ppdata}, based on which we define the choice stage constraint $T_d$ for each destination $d$ as\footnote{The choice stage constraint $T$ can be defined for each OD pair if we replace $N_d$ of (\ref{eq:T}) by $N_{od}$, the set of observations between $(o,d)$. However, such a definition requires more observations to have some for each OD pair and more computational effort, and thus we defined destination-specific constraints for this application.}
\begin{equation}
    \label{eq:T}
    T_d \equiv \max_{n \in N_d} \left[ \max\{\gamma D^d(o_n), J_n\} \right]
\end{equation}
where $N_d$ is the set of observations for $d$, and $o_n$ is the observed origin for observation $n$. We use the 75 percentile value ($\approx 1.34 = 4/3$, see Table \ref{tb:detour_stats}) for the detour rate $\gamma$ multiplied with the minimum number of steps $D^d(o)$ between $(o,d)$. 
We compared it with the number of steps observed $J_d$ for each destination and finally took the largest as $T_d$ so that all observations satisfy the prism constraint. 

We consider the following two specifications of the utility function:
\begin{subequations} \label{eq:pputil}
\begin{align}\centering
	&v(a|k) =   &\beta_{\rm{len}} {\rm Length}_a  &+ \beta_{\rm{cross}} {\rm Crosswalk}_a - 10 {\rm Uturn}_{a|k},\label{eq:pputil1}\\
	&v(a|k) =  &(\beta_{\rm{len}} +  \beta_{\rm{green}} {\rm Green}_a) {\rm Length}_a  &+ \beta_{\rm{cross}} {\rm Crosswalk}_a - 10 {\rm Uturn}_{a|k},\label{eq:pputil2}
\end{align}
\end{subequations}
where ${\rm Length}_a$ is the length (m/10) of link $a$, ${\rm Crosswalk}_a$ is the dummy variable of $a$ being a crosswalk, and ${\rm Green}_a$ is the dummy variable of green presence on link $a$\footnote{If a street $a$ has plants or trees on it or is along a park where some green is visible, ${\rm Green}_a$ is considered to be one. Otherwise, it is zero.} whose effect is captured by an interaction with ${\rm Length}_a$. Like in the experiments in the previous section, we fixed the coefficient of the uturn dummy variable to $-10$ and estimate the other coefficients $\beta_{\rm{len}}$, $\beta_{\rm{cross}}$, and $\beta_{\rm{green}}$. 
We expect that the link length and crosswalks have negative utility effects, while the green existence has a positive effect on pedestrian route choice behavior. 

Furthermore, we test an application to the NRL model \citep{Mai2015NRL} by considering the scale parameter $\mu$ to be link-specific $\mu \equiv \{\mu^d_k\}_{k \in \mathcal{A}, d \in \mathcal{D}}$, while it is normalized and fixed to one for the RL model. Specifically, we define $\mu^d_k = e^{\lambda^d_k}$ to impose a constraint on the scale parameter $\mu^d_k > 0$ \citep{Mai2015NRL}, and also define $\lambda^d_k = \omega \sqrt{{\rm SP}_{kd}}$ as a function of the shortest path length ${\rm SP}_{kd}$ (m/10) between link $k$ and destination $d$ to capture the increase/decrease in variance as the link approaches the destination \citep[e.g.,][]{Papola2013NGEV, oyama2022markovian}. The coefficient $\omega$ is the additional single parameter to be estimated.

We first compare the estimation results for specifications (\ref{eq:pputil1}) and (\ref{eq:pputil2}) in terms of model feasibility as well as interpretation.
Then, the goodness of fit and prediction performance of the models are compared. Finally, we investigate the impact of the choice stage constraint $T$ on model performance.

\subsubsection{Estimation Results}
Table \ref{tb:realestimation} reports the estimation results of the RL and Prism-RL models. First, for the specification (\ref{eq:pputil1}), we succeeded in estimating both the RL and Prism-RL models. Their estimates show similar mechanisms of pedestrian route choice behavior and are significantly different from zero at the 5\% significance level. From estimates signs, we found that pedestrians tend to walk paths with shorter lengths and avoid crosswalks, which meets our expectations. 

Then we compare the estimation results of the two specifications (\ref{eq:pputil1}) and (\ref{eq:pputil2}). For (\ref{eq:pputil2}), which adds an interaction term between the presence of green and the link length to (\ref{eq:pputil1}), we did not succeed in estimating the RL model although we have tried many initial points\footnote{We tried initial points including those with negative and large magnitude; specifically, we tested $1 \le \alpha_1 \le 20$ and $\alpha_2 \in \{0, -0.25\alpha_1\}$ for $\mathbold{\beta} = (-0.25 \alpha_1, -0.25 \alpha_1, \alpha_2)$, and also used the parameter estimates of the specification (\ref{eq:pputil1}). That is, 41 different initial points were tested in total, and the same for the NRL model.}. %
In contrast, the Prism-RL model did not experience the numerical problem, and we obtained the estimation result (the fifth column of Table \ref{tb:realestimation}). Like the experiment in Section \ref{sec:stability}, the estimation result of the Prism-RL model did not depend on the initial point. All estimates including $\hat{\beta}_{\rm green}$ were significantly different from zero at the 5\% significance level. Most importantly, the coefficient $\hat{\beta}_{\rm green}$ of green presence was estimated with a positive sign, while the other two coefficients remained negative. This result indicates that pedestrians are willing to walk streets with visible green, which meets our expectation and is consistent with the findings in the literature \citep[e.g.,][]{basu2022street}. 


\begin{table}[htb]
	\centering 
	\footnotesize
	\caption{Estimation results of the RL and Prism-RL models with the real observations. The second and third columns show the results for specification (\ref{eq:pputil1}), and the fourth and fifth columns for (\ref{eq:pputil2}). For specification (\ref{eq:pputil2}), the estimation of the RL model failed with all arbitrarily tested initial values; the reported result was obtained by the two-phase estimation procedure in Section \ref{sec:two-step}.}
	\label{tb:realestimation}
	\begin{tabular*}{\hsize}{@{\extracolsep{\fill}}lrrrr@{}}
		\toprule
		 & RL (\ref{eq:pputil1}) & Prism-RL (\ref{eq:pputil1}) 
		 & RL (\ref{eq:pputil2})$^{\dagger}$ & Prism-RL (\ref{eq:pputil2}) \\
		\midrule
        $\hat{\beta}_{\rm len}$ & \textbf{-0.297} & \textbf{-0.245} & \textbf{-0.318} & \textbf{-0.266} \\
        std.err. & 0.008 & 0.007 & 0.012 & 0.020 \\
        t-test & -38.832 & -37.264 & -25.523 & -13.283 \\
        $\hat{\beta}_{\rm cross}$ & \textbf{-0.924} & \textbf{-0.774} & \textbf{-0.936} & \textbf{-0.791} \\
        std.err. & 0.075 & 0.171 & 0.154 & 0.068 \\
        t-test & -12.237 & -4.517 & -6.082 & -11.638 \\
        $\hat{\beta}_{\rm green}$ &  &  & \textbf{0.054} & \textbf{0.049} \\
        std.err. &  &  & 0.014 & 0.010 \\
        t-test &  &  & 3.904 & 4.817 \\
        \midrule
        LL & -1772.972 & -1637.484 & -1743.794 & -1612.894 \\
        \#paths & 410 & 410 & 410 & 410 \\ 
		\bottomrule
		\multicolumn{5}{l}{$\dagger$: Estimated by the two-phase estimation procedure.}
	\end{tabular*}
\end{table}

Table \ref{tb:NRLestimation} reports the estimation results of the nested models. Unlike the RL case, the NRL model was successfully estimated for specification (\ref{eq:pputil2}). However, its success/failure depended on the selection of a starting point. The estimation of the NRL model (\ref{eq:pputil2}) failed for many of the initial parameters tested, including those of large magnitude and the estimates of the NRL model (\ref{eq:pputil1}). In contrast, the Prism-NRL model was estimated regardless of initial parameter values, which indicates that the NRL model benefits from the prism-based approach, as well as the RL model.
Estimates of $\hat{\beta}_{\rm len}$, $\hat{\beta}_{\rm cross}$ and $\hat{\beta}_{\rm green}$ remained statistically significantly different from zero at the 5\% significance level. The parameter $\hat{\omega}$ for the destination- and link-specific scales was also statistically significantly different from zero at the 5\% significance level. Given that the NRL models collapse to the RL models when $\omega$ is zero, the results captured well the underlying correlation among the utilities. Moreover, the positive sign of $\hat{\omega}$ implies that the scale $\mu^d_k$ and hence the variance become smaller as the location of the link approaches the destination, which we consider reasonable.

\begin{table}[htb]
	\centering 
	\footnotesize
	\caption{Estimation results of the NRL and Prism-NRL models with the real observations.}
	\label{tb:NRLestimation}
	\begin{tabular*}{\hsize}{@{\extracolsep{\fill}}lrrrr@{}}
		\toprule
		 & NRL (\ref{eq:pputil1}) & Prism-NRL (\ref{eq:pputil1}) 
		 & NRL (\ref{eq:pputil2}) & Prism-NRL (\ref{eq:pputil2}) \\
		\midrule
        $\hat{\beta}_{\rm len}$ & \textbf{-0.460} & \textbf{-0.445} & \textbf{-0.485} & \textbf{-0.469} \\
        std.err. & 0.030 & 0.039 & 0.082 & 0.062 \\
        t-test & -15.166 & -11.304 & -5.945 & -7.568 \\
        $\hat{\beta}_{\rm cross}$ & \textbf{-1.262} & \textbf{-1.206} & \textbf{-1.281} & \textbf{-1.206} \\
        std.err. & 0.163 & 0.120 & 0.280 & 0.201 \\
        t-test & -7.728 & -10.021 & -4.567 & -5.993 \\
        $\hat{\beta}_{\rm green}$ & - & - & \textbf{0.078} & \textbf{0.082} \\
        std.err. & - & - & 0.021 & 0.014 \\
        t-test & - & - & 3.690 & 5.855 \\
        $\hat{\omega}$ & \textbf{0.064} & \textbf{0.095} & \textbf{0.063} & \textbf{0.091} \\
        std.err. & 0.006 & 0.013 & 0.012 & 0.010 \\
        t-test & 9.942 & 7.402 & 5.459 & 8.769 \\
        \midrule
        LL & -1734.622 & -1587.079 & -1707.068 & -1565.531 \\
        \#paths & 410 & 410 & 410 & 410 \\ 
		\bottomrule
	\end{tabular*}
\end{table}


\subsubsection{Two-Phase Estimation}\label{sec:two-step}
With a positive network attribute, we failed to estimate the RL model with all the initial parameter values tested. The NRL model also suffered from the selection of a starting point, although it was successfully estimated with several ones. In general, it is difficult to find a good starting point that would allow us to avoid the numerical problem during the estimation of RL models. The prism-based approach can also contribute to solving this problem of initial parameter selection, where the Prism-RL model is viewed as an approximation of the RL model to provide a good starting point.

To this end, we introduce a two-phase estimation procedure. In the first phase, we estimate a Prism-RL model and obtain the estimation result. The estimates of the Prism-RL model are used in the second phase as a starting point for the estimation of the original RL model with the same utility specification. The RL model estimated in the second phase is considered a true model. As we have presented in the numerical results, the Prism-RL model can be estimated regardless of initial parameter values, and its estimates are good approximations of those of the RL model (i.e., their values are close to each other). If the true parameter value is a feasible solution for the value function of the RL models, a good starting point should mitigate the numerical issue during the estimation. 

We applied this two-phase estimation procedure to the RL model with specification (\ref{eq:pputil2}), using the estimates of the Prism-RL model (\ref{eq:pputil2}).
As a result, the RL model (\ref{eq:pputil2}) was successfully estimated and the estimation result is reported in the fourth column of Table \ref{tb:realestimation}. Again, the parameters were statistically and significantly estimated with the same signs as those of the Prism-RL model (\ref{eq:pputil2}), and the positive effect of green presence was also captured. This result shows the utility of the prism-based approach in the estimation of the original RL models in terms of solving the problem of initial parameter selection to mitigate the numerical issue of the value function during estimation.


\subsubsection{Model Comparison}\label{sec:comparison}
Next, we compare the models in terms of goodness of fit, and the summary is reported in Table \ref{tb:comparison}. As expected, the inclusion of the positive green attribute and the scale parameters improved the model performance. The likelihood ratio tests also showed the statistical preference of Model 4 to Models 1--3, and of Model 8 to Models 5--7, with the 95\% confidence level. Moreover, the prism-constrained models always obtained better log-likelihood and AIC values than the RL and NRL models. As a result, the best model in terms of in-sample fit seems to be Model 8, the Prism-NRL model with the positive attribute of green presence.

\begin{table}[htb]
	\centering 
	\footnotesize
	\caption{Model comparison. Likelihood ratio tests were performed to test statistical preference of Model 4 over Models 1--3, and of Model 8 over Models 5--7. Each pair of tested models assumes the same path set.}
	\label{tb:comparison}
	\begin{tabular*}{\hsize}{@{\extracolsep{\fill}}lcccccccc@{}}
		\toprule
		& \multicolumn{8}{c}{Model number}\\ \cmidrule(lr){2-9}
		& \#1 & \#2 & \#3 & \#4 & \#5 & \#6 & \#7 & \#8 \\
		\midrule
		Model & RL & NRL & RL & NRL & Prism-RL & Prism-NRL & Prism-RL & Prism-NRL \\
		$v(a|k)$ & (\ref{eq:pputil1}) & (\ref{eq:pputil1}) & (\ref{eq:pputil2}) & (\ref{eq:pputil2}) & (\ref{eq:pputil1}) & (\ref{eq:pputil1}) & (\ref{eq:pputil2}) & (\ref{eq:pputil2})\\
        Path set & Universal & Universal & Universal & Universal & Prism & Prism & Prism & Prism \\
        \#params & 2 & 3 & 3 & 4 & 2 & 3 & 3 & 4 \\
        $LL$ & -1772.97 & -1734.62 & -1743.794 & -1707.07 & -1637.48 & -1587.08 & -1612.89 & -1565.53 \\
        AIC & 3549.94 & 3475.24 & 3493.59& 3422.14 & 3278.97 & 3180.16 & 3231.79 & 3139.06 \\
        \midrule
        \multicolumn{3}{l}{\hspace{-2.2mm}Likelihood ratio test}  &  &  &  &  &  \\
        w.r.t. & \#4 & \#4 & \#4 & - & \#8 & \#8 & \#8 & - \\
        df & 2 & 1 & 1 & - & 2 & 1 & 1 & -\\
        $\chi^2$ & 131.80 & 55.10 & 73.45 & - & 143.91 & 43.10 & 94.73 & - \\
        p-value & <0.01 & <0.01 & <0.01 & - & <0.01 & <0.01 & <0.01 & - \\
		\bottomrule
	\end{tabular*}
\end{table}

To further investigate the difference in model performance on out-of-sample prediction, we performed a cross-validation. We randomly split the observations into estimation and holdout (validation) samples with a ratio of 80\% and 20\% and prepared 10 sets of them. The choice stage constraint $T$ was consistent with the case with all observations (\ref{eq:T}). The two-phase estimation procedure in Section \ref{sec:two-step} was used for the estimation of RL model (\ref{eq:pputil2}) for each sample.
The model performance was evaluated based on the log-likelihood obtained by applying the estimated model to the holdout sample: similarly to \cite{Mai2015NRL}, we computed the validation log-likelihood divided by the number of paths $LL_{i} = LL(\hat{\mathbold{\beta}}_{i}; \sigma_{i})/N_{i}$ for each holdout sample $i$ and then computed its average over samples $\overline{LL}_{i} = \frac{1}{p} \sum_{i=1}^{p} LL_{i}$, $\forall p \in \{1, \ldots, 10\}$. 

\begin{figure}[htb]
	\begin{center}
        \includegraphics[width=15cm]{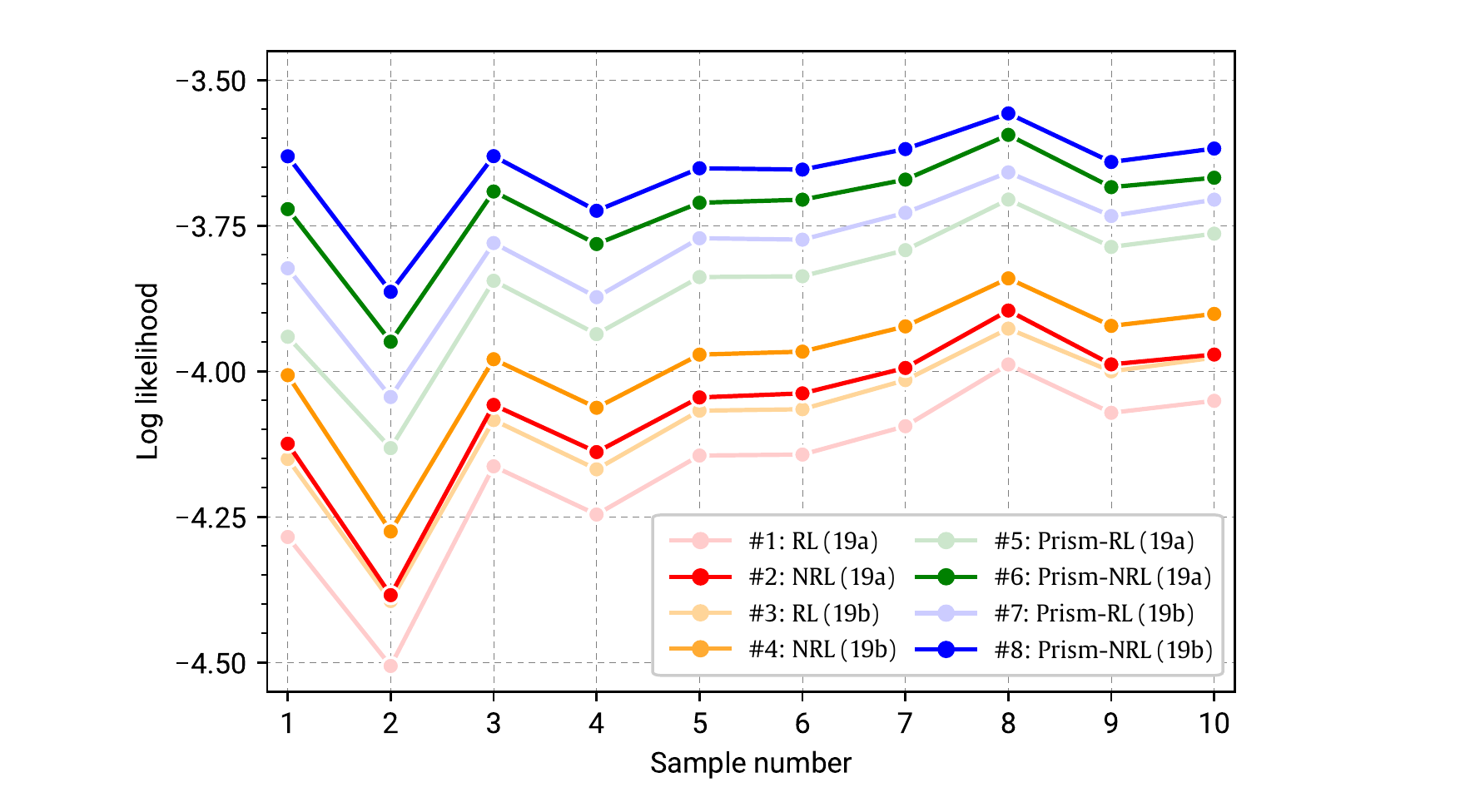}
		\caption{Validation results. The larger values (the upper positions) indicate the better model prediction performance. The model numbers correspond to those in Table \ref{tb:comparison}. A light-dark pair of the same color indicates the pair of RL-NRL models: Models 1-2 (red), Model 3-4 (orange), Models 5-6 (green), and Models 7-8 (blue). The plots colored in green and blue are the results of the models with the prism-constrained path set, while those in red are of the models without.}
		\label{fig:validation} 
	\end{center}
\end{figure}

Figure \ref{fig:validation} shows the validation results, and Table \ref{tb:validation} reports the average of the validation log-likelihood values over 10 holdout samples $\overline{LL}$ ($=\overline{LL}_{10}$). 
As general observations, the NRL-based models (Models 2,4,6,8) had higher prediction performance than the RL-based models (Models 1,3,5,7), and the inclusion of positive attributes also improved model prediction performance (Models 3,4,7,8). 


Notably, the prism-based approach made a significant difference in model prediction performance, demonstrated by the better validation log-likelihood values of Models 5-8 than Models 1-4 (Figure \ref{fig:validation}, Table \ref{tb:validation}). 
This difference in model prediction performance between the Prism-RL/Prism-NRL models and the RL/NRL models comes from the path sets assumed by those models. The RL/NRL models implicitly consider the universal path set, which includes infinite cyclic paths, while the Prism-RL/Prism-NRL models restrict such unrealistic paths by the prism constraint. This validation result suggests that the prism-based approach is not only a way of solving the numerical issue regarding the value function of the original RL models, but is also a more realistic description of route choice behavior. 


 
\begin{table}[htb]
	\centering 
	\footnotesize
	\caption{Average of validation log-likelihood values over 10 holdout samples. The larger values indicate the better model prediction performance. The model numbers correspond to those in Table \ref{tb:comparison}.}
	\label{tb:validation}
	\begin{tabular*}{\hsize}{@{\extracolsep{\fill}}lcccccccc@{}}
		\toprule
		& \#1 & \#2 & \#3 & \#4 & \#5 & \#6 & \#7 & \#8 \\
		\midrule
		$\overline{LL}$ &-4.051	&-3.971 &-3.977 &-3.902	&-3.764	&-3.668	&-3.705	&-3.618\\
		\bottomrule
	\end{tabular*}
\end{table}

\subsubsection{Impact of Choice Stage Constraint}\label{sec:Texp}
Finally, to explore the impact of the choice stage constraint $T$ that defines the prism on the model performance, we here test three additional values of detour rate $\gamma$ in (\ref{eq:T}): $1.25 (=5/4)$, $1.50 (=3/2)$, and $2.00 (=2/1)$. 
We tested both the Prism-RL and Prism-NRL models but focused on the specification (\ref{eq:pputil2}) that includes the interaction term of the green presence and link length. It is important to note first that the estimation of the Prism-RL models remained successful even when $T$ was large.

The parameter estimates and the final log-likelihood values for different prism constraints are reported in Table \ref{tb:difT}. 
For both models, the signs and scales of the estimates remained unchanged with different values of $\gamma$: $\hat{\beta}_{\rm len}$ and $\hat{\beta}_{\rm cross}$ were negative, $\hat{\beta}_{\rm green}$ was positive, and $\hat{\omega}$ ranged from $0.088$ to $0.092$. However, the ratio of the parameters $\hat{\beta}_{\rm len}$ and $\hat{\beta}_{\rm cross}$ of negative attributes systematically increased with increasing $\gamma$, for which an explanation is that the model adjusted the parameters to keep little probability of detour/cyclic paths included in the path set when $\gamma$ is large.
In fact, the smaller $\gamma$ values the tighter the prism constraint is, and the less paths are included in the path set of the models. As a result, behaviorally unrealistic paths are excluded, and the models with smaller values of $\gamma$ fit better in terms of log-likelihood, for both cases of the Prism-RL and Prism-NRL models. 
It should be noted that the fact that the estimation results depend on the value of $T$ means that the consistency property is not retained for the Prism-RL model.

\begin{table}[htb]
	\centering 
	\footnotesize
	\caption{Parameter estimates and final log-likelihood values of Prism-RL and Prism-NRL models with different prism constraints.}
	\label{tb:difT}
	\begin{tabular*}{\hsize}{@{\extracolsep{\fill}}lcccccc@{}}
		\toprule
		 & $\gamma$ & $\hat{\beta}_{\rm len}$ & $\hat{\beta}_{\rm cross}$ & $\hat{\beta}_{\rm green}$ & $\hat{\omega}$ & $LL$ \\
		\midrule
        Prism-RL (\ref{eq:pputil2}) & 1.25 & -0.265 & -0.785 & 0.049 & - & -1605.104\\
         & 1.34 & -0.266 & -0.791 & 0.049 & - & -1612.894 \\
         & 1.50 & -0.271 & -0.796 & 0.050 & - & -1632.277 \\
         & 2.00 & -0.284 & -0.817 & 0.052 & - & -1661.690 \\
        \midrule
        Prism-NRL (\ref{eq:pputil2}) & 1.25 & -0.464 & -1.191 & 0.082 & 0.090 & -1558.362 \\
         & 1.34 & -0.469 & -1.206 & 0.082 & 0.091 & -1565.531 \\
         & 1.50 & -0.482 & -1.227 & 0.085 & 0.092 & -1583.507 \\
         & 2.00 & -0.493 & -1.247 & 0.085 & 0.088 & -1613.565 \\
		\bottomrule
	\end{tabular*}
\end{table}


Although smaller values of $\gamma$ allow the models to better fit, defining a tightly constrained path set may have a problem for out-of-sample prediction; in other words, some paths that need to be predicted may not be contained in the path set (i.e., prism) defined in a data-oriented manner. We investigated this problem with the 10 holdout samples used for model comparison in Section \ref{sec:comparison}. In this validation, we define $T$ by (\ref{eq:T}) based only on the estimation sample. We also introduce the minimum choice stage constraint $\underline{T}$ and take the maximum between $T_d$ and $\underline{T}$ for each $d$. If the holdout sample contains a destination $d$ unobserved in the estimation sample, then the choice stage constraint for $d$ is $T_d = \underline{T}$. Then different values of $\gamma$ and $\underline{T}$ were tested. We counted the number of paths that are of the holdout sample and outside the prism defined based upon the estimation sample, which are reported in Table \ref{tb:ex_paths}. 
With small values of $\gamma$ and $\underline{T}$, the ratio of out-of-prism paths in the holdout sample was high. At maximum, when $\gamma = 1.25$ and $\underline{T} = 10$, 22 \% of paths did not satisfy the prism constraint defined by the estimation sample. The ratio decreased as the increase in $\gamma$ and $\underline{T}$. When $\underline{T} = 40$, for all holdout samples, all paths were contained in the prism defined based on the estimation sample. 
Therefore as expected, the tighter the prism, the more observed paths may not be feasible with respect to the prism constraint, while tight prism constraints have better numerical properties, such as fit in observations and computational efficiency.
To predict and compute the validation likelihood for such out-of-prism paths, we have to adjust $T_d$ and redefine prisms for corresponding destinations, which causes the inconsistency of the path set between estimation and prediction. 



\begin{table}[htb]
	\centering 
	\footnotesize
	\caption{Ratio of paths out of prism defined by the estimation sample for different $\gamma$ and $\underline{T}$: statistics over 10 holdout samples (mean [min, max]).}
	\label{tb:ex_paths}
	\begin{tabular*}{\hsize}{@{\extracolsep{\fill}}ccccc@{}}
		\toprule
		&\multicolumn{4}{c}{$\underline{T}$} \\\cmidrule(lr){2-5}
        $\gamma$ & 10 & 20 & 30 & 40 \\
        \midrule
        1.25 & 0.22 [0.13, 0.27] & 0.06 [0.02, 0.10] & 0.02 [0.00, 0.04] & 0.00 [0.00, 0.00] \\
        1.34 & 0.20 [0.12, 0.26] & 0.05 [0.02, 0.09] & 0.02 [0.00, 0.04] & 0.00 [0.00, 0.00] \\
        1.50 & 0.19 [0.11. 0.24] & 0.05 [0.02, 0.09] & 0.02 [0.00, 0.04] & 0.00 [0.00, 0.00] \\
        2.00 & 0.16 [0.07, 0.21] & 0.04 [0.00, 0.04] & 0.01 [0.00, 0.04] & 0.00 [0.00, 0.00] \\
		\bottomrule
	\end{tabular*}
\end{table}



\subsection{Discussion and Remarks}
This section presented a set of numerical results, which validated the advantages of the prism-based approach in model estimation and clarified its trade-off with limitations.
The Prism-RL model solved the numerical issue of the RL model and succeeded in capturing positive network attributes in the utility function while retaining implicit path enumeration. In the real application, with the same specification of the utility function, the prism-based models got a higher goodness of fit than the original RL models. This result implies that the prism-based approach not only solves the numerical problem of the RL models, but it may also provide a more realistic description of route choice behavior by restricting unrealistic paths by the prism constraint. 
On the other hand, the results also showed limitations of the prism-based approach. The parameter estimates depended on the choice of hyperparameter $T$; in other words, the Prism-RL models do not retain the consistency of the estimator, which is an important property of the original RL models. Although introducing a prism constraint improves numerical properties in model estimation, it may also restrict paths that are observed or need to be predicted. That is, the prism-based approach shares a common issue of inconsistency with approaches of sampling alternatives \citep[e.g.,][]{frejinger2009sampling, guevara2013sampling}, and thus there is a trade-off between its benefits and limitations.

In addition, a two-phase estimation procedure was introduced and tested for the estimation of the RL model with a positive attribute. It worked well and we successfully obtained the estimation result of the RL model. Note that with the estimates of the RL model (\ref{eq:pputil2}), the condition for a contraction mapping for the RL model \citep[in Theorem 1 of][]{mai2022undiscounted} did not hold: $94.2$\% of links satisfy $\sum_{a \in L} M_{ka} < 1$ and $\max_{k \in L} \sum_{a \in L} M_{ka} = 1.503$. However, the fact that the Bellman equation is not a contraction does not mean that a solution does not exist. Although we could solve the value function with the estimates obtained by using the two-phase estimation procedure, \cite{mai2022undiscounted} proposed algorithms to solve the value function that work better than the simple value iteration. Integration of such algorithms within the two-phase estimation may improve the numerical property and be particularly useful in cases where the Bellman equation is not a contraction, which we consider in future work.

The numerical examples used the original topology of the network to define the choice stage, which implies that we restrict the feasible path set by the maximum number of links. According to the definition, we used the detour rate that is based on the number of links contained in the path. However, by changing the definition of the choice stage, the Prism-RL model can take another type of constraint, such as the maximum length or time of paths \citep[see][for the detail]{oyama2019prism}.
The prism constraint defined based upon a specific variable such as travel distance or time may be easier to behaviorally interpret, but they require network editing (i.e., link splitting into unit length), which may increase the number of states and makes the computation inefficient. 
In contrast, the prism constraint based on the number of links, which this study used, does not depend on a specific variable or require network editing. Furthermore, when the network has cycles consisting of very short links, the constraint of the maximum link number in a path should be able to restrict such cyclic route choice behavior.

Note that even though the Prism-RL model is executable in a feasible time, it requires more computational effort than the RL model, and the computational time depends on the definition of choice stage constraint. This is because the linear system (\ref{eq:linearsystem}) of the RL model, which can be efficiently computed, is not available for the Prism-RL model. However, an advantage of (\ref{eq:linearsystem-p}) is its independence of whether the model is linear or nonlinear, and in fact, the scale of the required estimation time for the Prism-NRL model remained unchanged from that for the Prism-RL model (see \ref{app:runtime}). 

\section{Application Potential}\label{sec:direction}\noindent
Given the findings from the present results, in this section we summarize and discuss two application directions of the prism-based approach for the readers to use it in future studies.

In the first direction, the Prism-RL model is considered as a final model, where the prism-based approach is used as an algorithm of sampling alternatives for link-based route choice modeling. 
In this case, the Prism-RL model can be interpreted as a different model from the RL model and describes nonidentical behavioral mechanisms of travelers in different choice stages (as discussed in Figure \ref{fig:prism}). Although the selection of the hyperparameter $T$ is still required, defining it in a data-oriented manner can restrict unrealistic paths and achieve a better fit and higher prediction performance of the model than the original RL models. More importantly, the Prism-RL model does not experience the numerical problem of the value function and thus can deal with more flexible utility functions than the RL model, such as including more number of and/or positive attributes.
This direction is particularly useful for applications, such as pedestrian route choice\footnote{In the present case study, Figure \ref{fig:kannai_paths} in \ref{app:ppdata} shows examples of diverse walking paths of pedestrians, suggesting a difficulty in the definition of the set of paths. Pedestrian route choices are also likely to be affected by some positive attributes of the streets.} or sequential destination choice behavior, where the path set definition is a difficult task due to the diversity of paths, and the evaluation of attractiveness (various positive attributes) of alternatives is needed. In such cases, classical approaches, i.e., route choice models with an explicit path enumeration algorithm \citep[for reviews, see e.g.,][]{prato2009route}, may not be very useful because it is difficult to include diverse paths in the choice set. The unrestricted path set of the RL model also involves the numerical problem of the value function when evaluating positive network attributes. The Prism-RL model addresses both of these challenges by implicitly restricting unrealistic paths, as shown in the real case study of a pedestrian network. 

However, the consistency of the maximum likelihood estimator is not retained for the Prism-RL model, and its estimates depend on the hyperparameter of the choice stage constraint. Since the consistency property is an important motivation behind the RL models, this is a main limitation of the Prism-RL model. The second direction of application therefore regards the RL model as a true model and estimates the Prism-RL model as its approximation providing a good starting point. That is, the RL model is estimated based on the two-phase estimation procedure described in Section \ref{sec:two-step}: estimate the Prism-RL model first and then use its estimates as a starting point for the estimation of the RL model with the same utility specification\footnote{The choice of hyperparameter $T$ in the first phase should not be a problem because the signs and scales of the estimates of the Prism-RL models remain unchanged across different $T$ as shown in Table \ref{tb:difT}. That is, with any $T$ their estimates work to some extent as a good starting point in the second phase.}. When the modeler needs to capture positive network attributes, the RL model is likely to suffer from the numerical problem of the value function during the parameter search by a nonlinear optimization algorithm. A very good starting point for the algorithm should mitigate this problem, but it is often difficult to find. The use of the Prism-RL model as an approximation addresses the selection problem of a starting point and solves the numerical issue if the true parameter value is a feasible solution for the value function of the RL model. This benefit was well demonstrated by the pedestrian case study in the previous section: the estimation of the RL model with a positive attribute failed with all arbitrarily tested initial values, but the two-phase estimation successfully obtained the result. Importantly, in this approach, the RL model estimated in the second phase retains the consistency property, and the RL estimator can be used for prediction with the prism-based approach so that the parameter does not depend on the choice set. In this case, the prediction with the prism-based approach has computational advantages such as efficiency and mitigation of cyclic flows \citep[see][]{oyama2019prism}.

\section{Conclusion}\label{sec:conclusion}\noindent
In this study, we conducted extensive numerical experiments to examine the properties of the Prism-RL model in parameter estimation. We obtained two main important results. First, the Prism-RL models were successfully estimated regardless of the initial and true parameter values, even in the cases where the RL models cannot be estimated due to the numerical issue of the value function during the parameter search process. In the real application, the prism-based approach successfully captured the positive effect of green presence on pedestrian route choice behavior. It has not often been possible to capture such positive attributes with the previous RL models due to the numerical issue. Second, the Prism-RL models achieved better fit and prediction than the RL models by restricting unrealistic paths such as largely detour or cyclic ones, which was well demonstrated by cross-validation. We presented a possibility to define the prism-based path set in a data-oriented manner, using the information of observed detour rates. This enabled a more realistic description of route choice behavior, compared to the RL models that assume all feasible paths including infinite cyclic paths in the choice set. 

This study also analyzed the impact of the determination of the choice stage constraint parameter $T$ on the estimation results. Although the prism-constrained path set allowed the model to better fit in observations, its definition affected the estimation result of the Prism-RL model, and the consistency property was not retained. To address this limitation, we presented a two-phase estimation procedure in which the RL model is considered the true model, and the Prism-RL model is estimated to provide a good starting point for the RL model. 

In summary, there are two directions for further applications of the prism-based approach, both of which contribute to capturing positive network attributes without path enumeration. The first direction estimates the Prism-RL model as a true model and may be able to deal with more flexible utility functions compared to the RL model, such as including more and positive network attributes or a positive interaction between agents. However, in this approach, the consistency property of the estimator is not retained for the Prism-RL model. The second direction therefore considers the RL model as a true model and estimates it by using the two-phase estimation procedure. The Prism-RL model is estimated in its first phase and provides a good starting point for the estimation of the RL model to mitigate the numerical problem of the value function. 

Around the world, urban design projects are recently planned to make better places in city centers where people visit for various activities and enjoy walking \citep[e.g.,][]{mueller2020changing}. To properly assess the impacts of such projects, it is not sufficient to capture only negative attributes of streets. It has been needed to develop a method that is able to analyze the attractiveness having positive effects on pedestrians' behavior while capturing the diversity of their route choice, and our proposed method can be considered as a significant contribution in such a context. 
Although this study focuses on the methodology and showed a pedestrian network application as an example to validate the presented model, including only a variable of green presence as a positive attribute is not sufficient to understand pedestrian route choice behavior. Recent studies have shown that, for example, wide sidewalks and a high number of available amenities as well as green presence positively influence pedestrian route choice \citep[e.g.,][]{sevtsuk2021big, basu2022street}. Because most previous studies used a classical route choice model that requires path enumeration, e.g., the path-size logit model \citep{Ben-Akiva1999}, it is an interesting topic for future research to use the Prism-RL model for a more detailed analysis on pedestrian route choice and compare it with the findings in the literature.

Having a way of solving the numerical issue of the RL model, we consider many open and potentially valuable research directions for further study. Recursive modeling of behavior is a general framework, and its application is not limited to route choice analysis. Modeling of trip chaining behavior \citep[e.g.,][]{kitamura1984incorporating}, i.e., a sequence of destination choices, is a particular application in which we need to deal with choice set and adequately evaluate the attractiveness of destinations. The Prism-RL model will be useful for such a case, where the choice stage constraint can be regarded as the maximum number of places that a traveler can visit and is more easily defined compared to route choice modeling, and this research direction seems appealing. 
Network design problems with positive impacts can be formulated by using the Prism-RL model as a demand-side simulation tool. Possible model extensions include the incorporation of uncertainty or travelers' heterogeneity in constraints, algorithm developments to speed up the estimation, and applications to traffic assignment models.

\section*{Acknowledgements}\noindent
This research was supported by JSPS KAKENHI Grant Number 20K14899, and the Committee on Advanced Road Technology, Ministry of Land, Infrastructure, Transport, and Tourism, Japan \#2020-1. The data for the case study was collected through a Probe Person survey, a complementary survey of the Sixth Tokyo Metropolitan Region Person Trip Survey. The author thanks Daisuke Fukuda for his valuable discussion on the research and is also grateful to three anonymous reviewers for their detailed comments on the earlier version of the manuscript. Their constructive feedback enabled a significant improvement of the quality of the manuscript.



\appendix

\section{Sioux Falls Network Data}\label{app:network}
\begin{figure}[htb]
	\begin{center}
		\includegraphics[width=15cm]{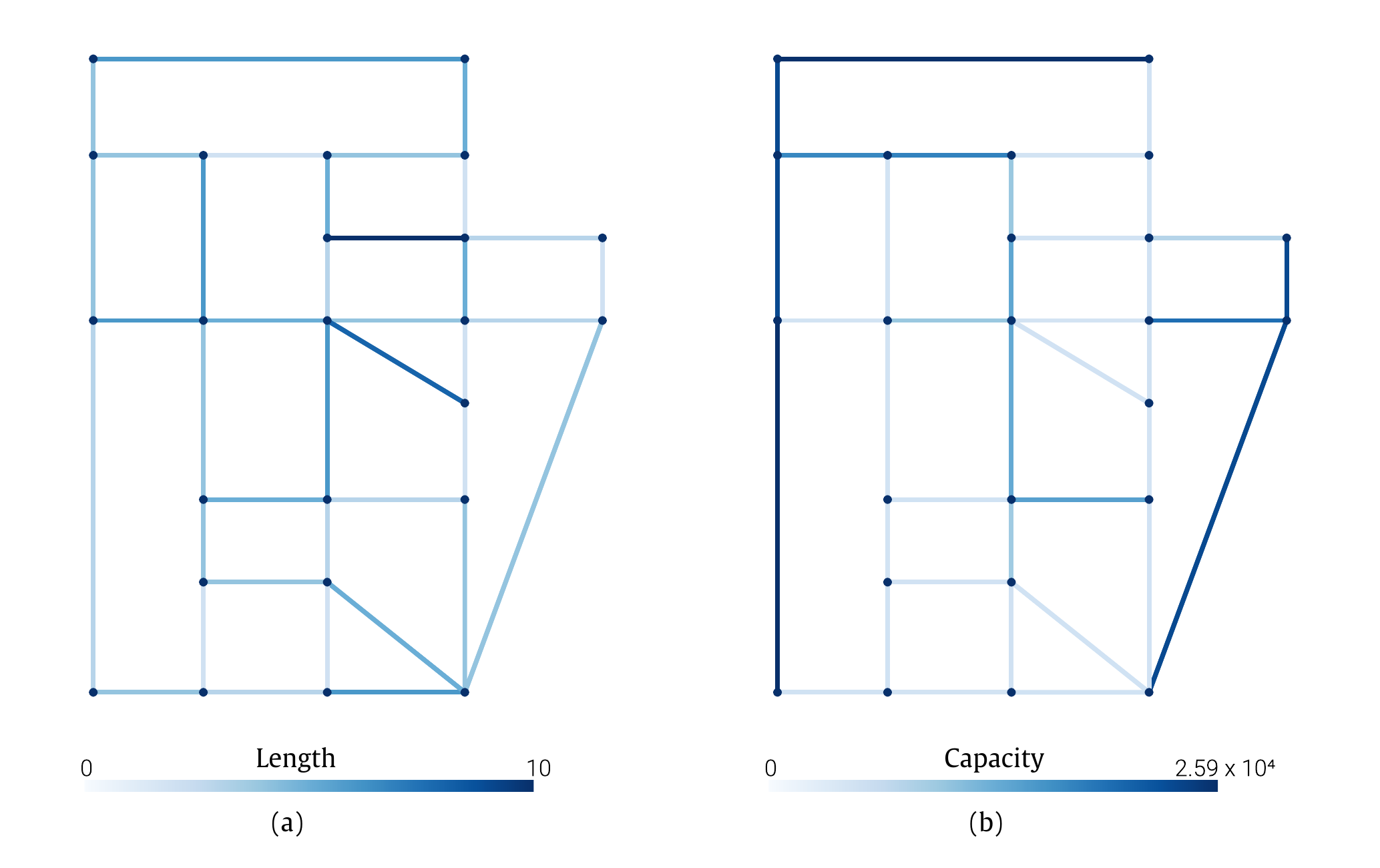}
		\caption{Link variables in the Sioux Falls network: (a) length, (b) capacity. Deep colored lines indicate links with larger magnitude of variables.}
		\label{fig:net} 
	\end{center}
\end{figure}

\noindent
Figure \ref{fig:net} shows the Sioux Falls network, where (a) and (b) map the values of link length and link capacity. Note that the link lengths are equal to the free-flow link costs in the provided data \citep{SiouxFalls2016}, which is why the length variable is not consistent with the visual length of Figure \ref{fig:net}.

\section{Impact of $T$ in Sioux Falls Experiment}\label{app:SF_difT}\noindent
To examine the impact of $T$ on the estimation results in the Sioux Falls experiment, we estimated the Prism-RL model with different $T$ values, using the full sample generated in Section \ref{sec:siouxfalls}. Tables \ref{tb:SF_difT1} and \ref{tb:SF_difT2} report the parameter estimates and final log-likelihood values, respectively for the cases of the true models with $(\tilde{\beta}_{\rm len}, \tilde{\beta}_{\rm cap}) = (-2.0, -1.5)$ and $(\tilde{\beta}_{\rm len}, \tilde{\beta}_{\rm cap}) = (-2.5, 2.0)$.
For both cases, the final log-likelihood values for $T=25, 50, 75, 100$ remained the same to the eighth decimal place. The parameter estimates for the different $T$ values were also consistent. The Prism-RL model reproduced the parameters and log-likelihood values very close to those of the true models.
As such, the setting of $T$ had little impact on the estimation in this Sioux Falls experiment. This is because we simulated the observations by using the RL model, which corresponds to the Prism-RL model with $T \rightarrow \infty$. Because the simulated observations did not contain any path with a loop or a large detour, even with a small value of $T$ the Prism-RL model successfully reproduced the parameters of the true model. 

\begin{table}[htb]
	\centering 
	\footnotesize
	\caption{Estimation results of the Sioux Falls experiment with different $T$ values: the case of $(\tilde{\beta}_{\rm len}, \tilde{\beta}_{\rm cap}) = (-2.0, -1.5)$.}
	\label{tb:SF_difT1}
	\begin{tabular*}{\hsize}{@{\extracolsep{\fill}}lcccc@{}}
		\toprule
        &$T$ & $\hat{\beta}_{\rm len}$ & $\hat{\beta}_{\rm cap}$ & $LL$ \\
        \midrule
        RL & - & -1.99970769 & -1.50474450 & -4378.74273614 \\
        Prism-RL & 10 & -1.99970765 & -1.50474443 & -4378.74273395 \\
         & 25 & -1.99970774 & -1.50474454 & -4378.74273614 \\
         & 50 & -1.99970774 & -1.50474454 & -4378.74273614 \\
         & 75 & -1.99970774 & -1.50474454 & -4378.74273614 \\
         & 100 & -1.99970774 & -1.50474454 & -4378.74273614 \\
         \midrule
         True RL model & - & -2.00 & -1.50 & -4378.85197614 \\ 
		\bottomrule
	\end{tabular*}
\end{table}

\begin{table}[htb]
	\centering 
	\footnotesize
	\caption{Estimation results of the Sioux Falls experiment with different $T$ values: the case of $(\tilde{\beta}_{\rm len}, \tilde{\beta}_{\rm cap}) = (-2.5, 2.0)$.}
	\label{tb:SF_difT2}
	\begin{tabular*}{\hsize}{@{\extracolsep{\fill}}lcccc@{}}
		\toprule
        &$T$ & $\hat{\beta}_{\rm len}$ & $\hat{\beta}_{\rm cap}$ & $LL$ \\
        \midrule
        RL$^{*}$      & - & -2.49716457 & 2.00055747 & -6729.71207206 \\
        Prism-RL & 10 & -2.49716340 & 2.00055637 & -6729.71127367 \\
                 & 25 & -2.49716488 & 2.00055776 & -6729.71207206 \\
                 & 50 & -2.49716488 & 2.00055773 & -6729.71207206 \\
                 & 75 & -2.49716497 & 2.00055780 & -6729.71207206 \\
                 & 100 & -2.49716499 & 2.00055779 & -6729.71207206 \\
         \midrule
         True RL model & - & -2.50 & 2.00 & -6730.22334234 \\ 
		\bottomrule
		\multicolumn{5}{l}{$^{*}$: Initial parameter values were set to $(-4, 3)$.}
	\end{tabular*}
\end{table}

Table \ref{tb:runtime_SF} reports the estimation times of the RL model and the Prism-RL model with different $T$ values. The reported times are the averages over 10 samples used in Section \ref{sec:reproducibility}.
The Prism-RL model required more computational time than the RL model, and its increase was approximately linear in the choice stage constraint $T$. The case of $(\tilde{\beta}_{\rm len}, \tilde{\beta}_{\rm cap}) = (-2.5, 2.0)$ took more time because it required more iterations to converge than the case of $(\tilde{\beta}_{\rm len}, \tilde{\beta}_{\rm cap}) = (-2.0, -1.5)$. 

\begin{table}[htb]
	\centering 
	\footnotesize
	\caption{Estimation time in seconds for the Sioux Falls network experiment. The reported values are the average times over 10 samples.}
	\label{tb:runtime_SF}
	\begin{tabular*}{\hsize}{@{\extracolsep{\fill}}lcccccc@{}}
		\toprule
		& & \multicolumn{5}{l}{Prism-RL}\\ \cmidrule(rl){3-7}
        $(\tilde{\beta}_{\rm len}, \tilde{\beta}_{\rm cap})$ & RL & $T = 10$ & $25$ & $50$ & $75$ & $100$ \\
        \midrule
        $(-2.0, -1.5)$ & 0.84$^{~}$ & 1.69 & 5.18 & 11.45 & 17.72 & 25.03 \\
        $(-2.5, 2.0)$ & 1.02$^{*}$ & 2.40 & 6.50 & 14.36 & 22.61 & 31.22 \\
		\bottomrule
		\multicolumn{7}{l}{$^{*}$: Initial parameter values were set to $(-4, 3)$.}
	\end{tabular*}
\end{table}

\section{Pedestrian Network and Route Choice Data}\label{app:ppdata}
\begin{figure}[htb]
	\begin{center}
		\includegraphics[width=15cm]{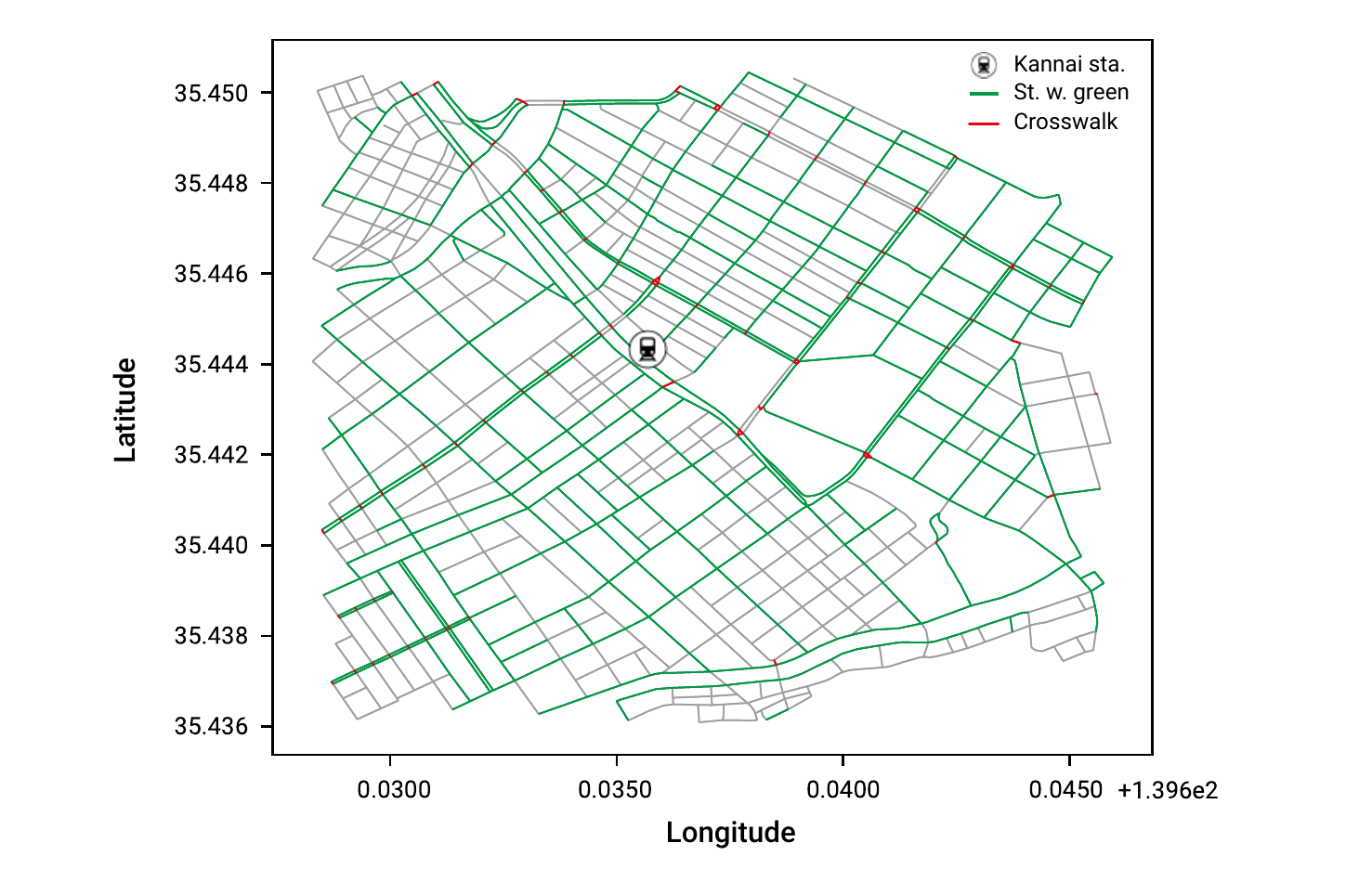}
		\caption{Pedestrian network for real application. The area is a mile square centered on the Kannai station. Green lines indicate streets with visible green, and red lines are crosswalks.}
		\label{fig:kannai} 
	\end{center}
\end{figure}

\noindent
Figure \ref{fig:kannai} shows the pedestrian network for the real application, which includes street data within a mile square centered on the Kannai station, Yokohama city, Japan. The dummy variables ${\rm Crosswalk}_a$ and ${\rm Green}_a$ are also mapped. The Kannai district has many streets with visible green.
Note that we obtained the original network data from \cite{OpenStreetMap} using the Python library OSMnx \citep{boeing2017osmnx}, on which we added the street variables.

\begin{figure}[htb]
	\begin{center}
		\includegraphics[width=15cm]{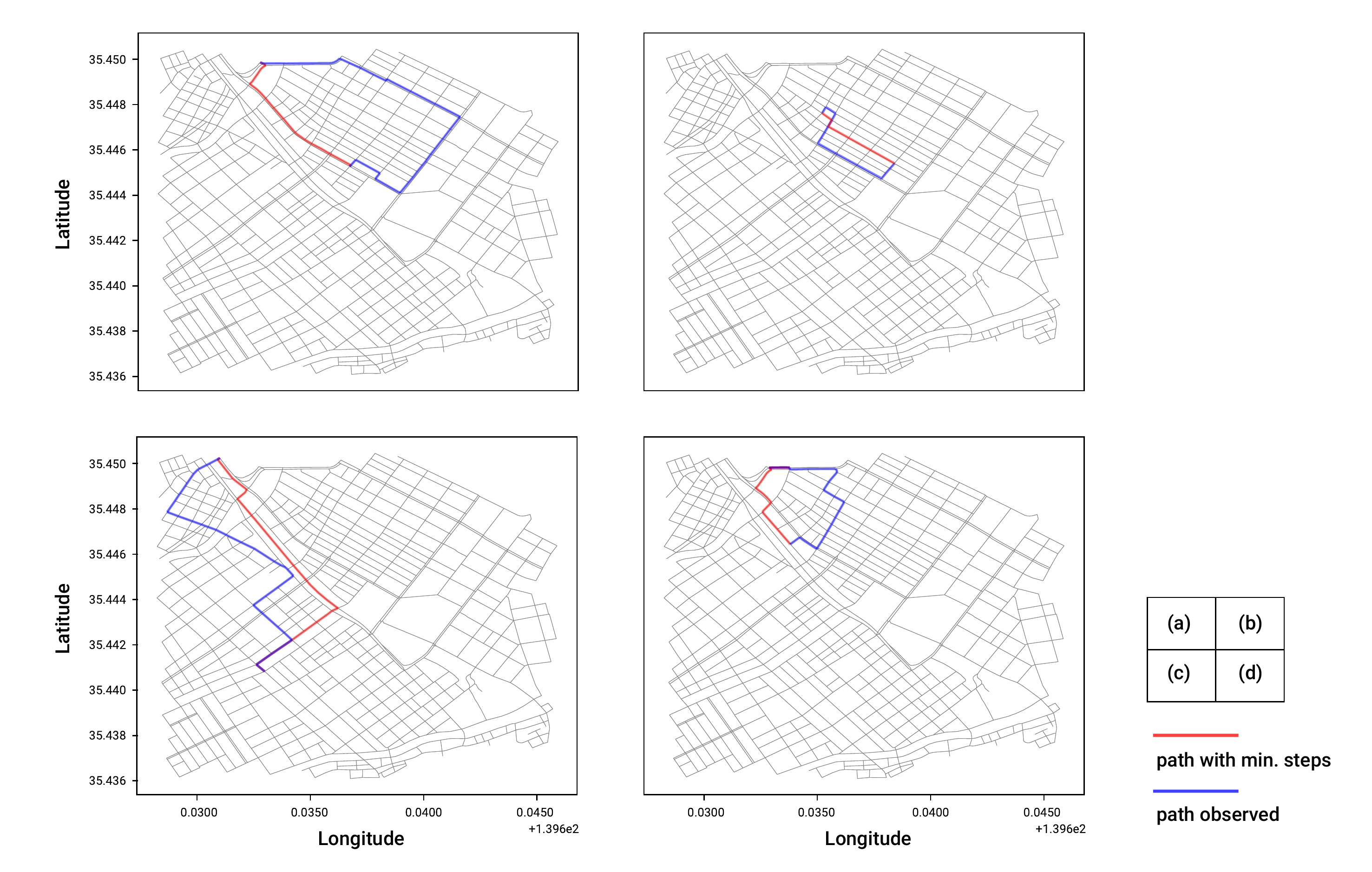}
		\caption{Examples of observed detour walking paths. Blue paths are the observed paths, and red paths are paths with minimum steps between the observed OD pairs.}
		\label{fig:kannai_paths} 
	\end{center}
\end{figure}

Figure \ref{fig:kannai_paths} shows four examples of observed walking paths with high detour rates. We find that pedestrians walk a variety of paths, sometimes taking a large detour. We also observe that, in the example (b), the upper right, the pedestrian chose a path that is longer than the shortest path but passes through streets with green presence.

\clearpage

\begin{figure}[htb]
	\begin{center}
		\includegraphics[width=15cm]{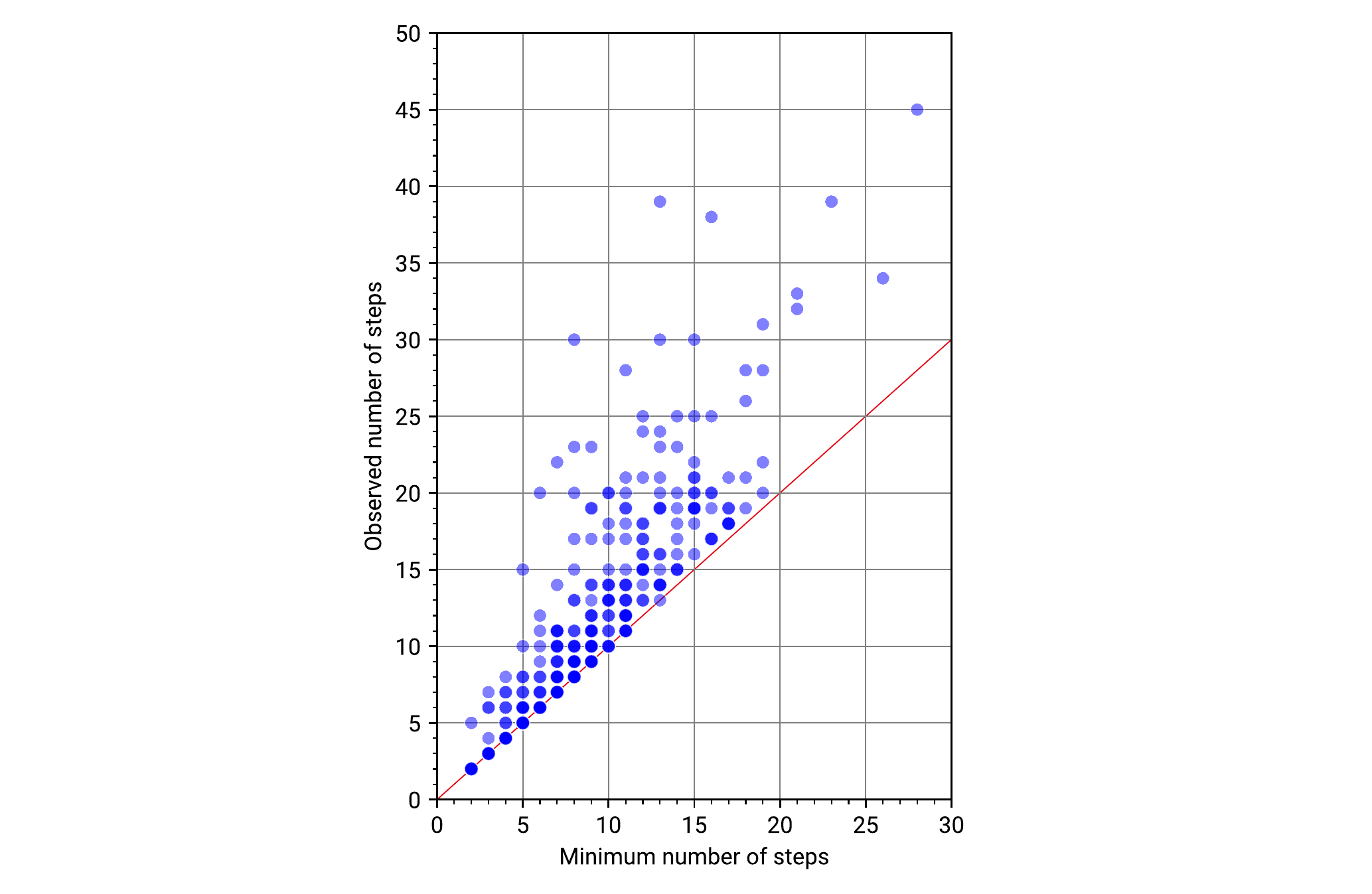}
		\caption{Plot of detour characteristics of the observed paths. The x-axis is the minimum step for each observed OD pair, and the y-axis is the steps actually used. Deep colors mean that the points are observed many times. The red line is the identity line.}
		\label{fig:detour} 
	\end{center}
\end{figure}

Figure \ref{fig:detour} is the plot of the detour characteristics of the observed walking trips, and Table \ref{tb:detour_stats} reports the statistics of the observed detour rates, where we define a detour rate as the number of observed steps (links) divided by the minimum number of steps. The detour rates of 75\% of the observed paths are below $4/3\approx 1.34$. We used this value as well as the maximum step number for the decision of the choice stage constraint $T_d$ for each destination $d$, as in Eq.(\ref{eq:T}).

\begin{table}[htb]
	\centering 
	\footnotesize
	\caption{Statistics of detour rates, or the observed number of steps divided by the minimum number of steps.}
	\label{tb:detour_stats}
	\begin{tabular*}{\hsize}{@{\extracolsep{\fill}}cccccccc@{}}
		\toprule
        Count & Mean & Std. & Min. & 25\% & 50\% & 75\% & Max. \\
        \midrule
        410 & 1.24 & 0.39 & 1.00 & 1.00 & 1.08 & 1.33 & 3.75 \\
		\bottomrule
	\end{tabular*}
\end{table}

\section{Computational Time for Estimation}\label{app:runtime}\noindent
Table \ref{tb:runtime_PP} reports the estimation times for the real application in Section \ref{sec:application}. For the estimation of the NRL and Prism-NRL models, we used the parameter estimates of the RL and Prism-RL models for the starting points.
The RL model was very fast because its value function can be efficiently computed by solving the linear system (\ref{eq:linearsystem}). The Prism-RL models required more computational effort than the RL model. However, the scale does not change between the Prism-RL and Prism-NRL models because their solution methods are the same, as described in Section \ref{sec:prism_value}. The fact that the computational complexity does not depend on the linearity of the model may be an advantage of the prism-based approach compared to the original RL models. Note that we used a fixed and high precision for the value iteration during the estimation of the NRL model, but it could be estimated faster by using the dynamic accuracy \citep{Mai2015NRL} and/or the algorithms proposed in \cite{mai2022undiscounted}.

\begin{table}[htb]
	\centering 
	\footnotesize
	\caption{Estimation time in seconds for the real application. The reported values are the average times over the 10 hold-out samples.}
	\label{tb:runtime_PP}
	\begin{tabular*}{\hsize}{@{\extracolsep{\fill}}lcccccccc@{}}
		\toprule
		&\multicolumn{2}{c}{RL} & \multicolumn{2}{c}{NRL}  & \multicolumn{2}{c}{Prism-RL}  & \multicolumn{2}{c}{Prism-NRL} \\ \cmidrule{2-9}
		$v(a|k)$&(\ref{eq:pputil1})&(\ref{eq:pputil2}) & (\ref{eq:pputil1}) & (\ref{eq:pputil2})& (\ref{eq:pputil1}) & (\ref{eq:pputil2})& (\ref{eq:pputil1}) & (\ref{eq:pputil2})\\
		\midrule
        CPU time&36.40 & 60.04 & 1096.82 & 1282.22 & 386.24 & 768.89 & 781.48 & 900.17 \\
		\bottomrule
	\end{tabular*}
\end{table}

\bibliographystyle{elsarticle-harv}
\bibliography{prism}

\end{document}